\documentclass[11pt,a4paper]{article}
\usepackage{jheppub}

\usepackage{amsmath}
\usepackage{amsfonts}
\usepackage{slashed}
\usepackage{graphicx}
\usepackage{mathrsfs}
\usepackage{tikz}
\usepackage[numbers,sort&compress]{natbib}

\makeatletter
\g@addto@macro\bfseries{\boldmath}
\makeatother

\newcommand{\eps}{\epsilon}
\newcommand{\ord}{\begin{cal}O\end{cal}}

\def\beq{\begin{equation}}
\def\eeq{\end{equation}}
\def\bsp#1\esp{\begin{split}#1\end{split}}

\makeatletter
\newcommand{\IEIF}{%
  \def\@IEIFsep{(}%
  I_F\@IEIFi
}
\newcommand\@IEIFi{\@ifnextchar\stopIEIF{\@IEIFend}{\@IEIFii}}
\newcommand\@IEIFii[4]{%
  \big\@IEIFsep
  \begin{smallmatrix}
    #1 & #2 \\
    #3 & #4
  \end{smallmatrix}
  \def\@IEIFsep{|}
  \@IEIFi
}
\newcommand\@IEIFend[2]{%
  ; #2 \bigr)
}
\makeatother

\newcommand{\cA}{\begin{cal}A\end{cal}}

\newcommand{\cE}{\begin{cal}E\end{cal}}

\newcommand{\cG}{\begin{cal}G\end{cal}}

\newcommand{\cN}{\begin{cal}N\end{cal}}

\newcommand{\cZ}{\begin{cal}Z\end{cal}}

\newcommand{\Efe}[4]{{\textrm{E}_4}\!\left(\begin{smallmatrix}#1\\#2\end{smallmatrix};#3,#4\right)}

\newcommand{\gamtt}[4]{{\widetilde{\Gamma}}\!\left(\begin{smallmatrix}#1\\#2\end{smallmatrix};#3,#4\right)}

\newcommand{\cEf}[4]{{\mathcal{E}_4}\!\left(\begin{smallmatrix}#1\\#2\end{smallmatrix};#3,#4\right)}

\renewcommand{\ln}{\log}

\allowdisplaybreaks

\title{Elliptic Feynman integrals and pure functions}

\author[a]{Johannes Broedel} 
\author[b,c]{Claude Duhr}
\author[d]{Falko Dulat}
\author[b]{Brenda Penante}
\author[b]{Lorenzo Tancredi}

\affiliation[a]{Institut f\"{u}r Mathematik und Institut f\"{u}r Physik,
Humboldt-Universit\"{a}t zu Berlin,\\
IRIS Adlershof, Zum Grossen Windkanal 6, 12489 Berlin, Germany} 
\affiliation[b]{Theoretical Physics Department, CERN, Geneva, Switzerland} 
\affiliation[c]{Center for Cosmology, Particle Physics and Phenomenology (CP3),\\
Universit\'e Catholique de Louvain, 1348 Louvain-La-Neuve, Belgium}
\affiliation[d]{SLAC National Accelerator Laboratory, Stanford University, Stanford, CA 94309, USA}

\emailAdd{jbroedel@physik.hu-berlin.de}
\emailAdd{claude.duhr@cern.ch}
\emailAdd{dulatf@slac.stanford.edu}
\emailAdd{b.penante@cern.ch}
\emailAdd{lorenzo.tancredi@cern.ch}

\abstract{
We propose a variant of elliptic multiple polylogarithms that have at most logarithmic singularities in all variables and satisfy a differential equation without homogeneous term. We investigate several non-trivial elliptic two-loop Feynman integrals with up to three external legs and express them in terms of our functions. We observe that in all cases they evaluate to pure combinations of elliptic multiple polylogarithms of uniform weight.
This is the first time that a notion of uniform weight is observed in the context of Feynman integrals that evaluate to elliptic polylogarithms.
}

\preprint{\begin{minipage}[t]{8cm}\begin{flushright}CP3-18-58, CERN-TH-2018-211\\
            HU-Mathematik-2018-09, HU-EP-18/29\\
        SLAC-PUB-17336\end{flushright}\end{minipage}}

\begin{document}

\maketitle

\catcode`\@=11
\font\manfnt=manfnt
\def\Watchout{\@ifnextchar [{\W@tchout}{\W@tchout[1]}}
\def\W@tchout[#1]{{\manfnt\@tempcnta#1\relax%
  \@whilenum\@tempcnta>\z@\do{%
    \char"7F\hskip 0.3em\advance\@tempcnta\m@ne}}}
\let\foo\W@tchout
\def\dubious{\@ifnextchar[{\@dubious}{\@dubious[1]}}
\let\enddubious\endlist
\def\@dubious[#1]{%
  \setbox\@tempboxa\hbox{\@W@tchout#1}
  \@tempdima\wd\@tempboxa
  \list{}{\leftmargin\@tempdima}\item[\hbox to 0pt{\hss\@W@tchout#1}]}
\def\@W@tchout#1{\W@tchout[#1]}
\catcode`\@=12


\section{Introduction}
\label{sec:intro}

In perturbation theory physical observables are expanded into a series in the coupling constants of the theory. The $n$-th order in perturbation theory involves a sum of $n$-loop Feynman diagrams with a fixed set of external legs that need to be integrated over the momentum flowing in each of the $n$ loops. Hence, the evaluation of higher orders in perturbative quantum field theory (QFT) is tightly connected to the computation of multi-loop Feynman integrals.
Over the last decade we have witnessed an enormous increase in our ability to compute Feynman integrals analytically. This progress can be traced back, among other things, to an improved understanding of multi-loop integrals and the mathematics underlying them.

Unitarity implies that Feynman integrals must have discontinuities, and therefore they must evaluate to special functions that reproduce this branch cut structure.
We have nowadays a rather solid and complete mathematical understanding of the simplest class of special functions, called \emph{multiple polylogarithms} (MPLs), that show up in multi-loop computations~\cite{Goncharov:1995,Remiddi:1999ew,Goncharov:1998kja,Vollinga:2004sn}. MPLs are not only relevant to precision computations in QFT, but they
are an active area of research also in contemporary pure mathematics. The insight into the mathematical structure of MPLs has ultimately led to the development of novel computational techniques for Feynman integrals, cf., e.g., refs.~\cite{Goncharov:2010jf,Goncharov:2009tja,ChenSymbol,Brown:2009qja,Duhr:2011zq,Ablinger:2011te,Buehler:2011ev,Duhr:2012fh,Anastasiou:2013srw,Henn:2013pwa,Ablinger:2013cf,Panzer:2014caa,Bogner:2014mha,Duhr:2014woa,Ablinger:2014bra,Henn:2014qga,Frellesvig:2016ske,Frellesvig:2018lmm}. Inherent to many of these techniques is the concept of \emph{pure functions} and \emph{pure integrals}~\cite{ArkaniHamed:2010gh}, defined as integrals such that all the non-vanishing residues of the integrand are equal up to a sign. MPLs are the prototypical examples of pure functions, and loosely speaking one may think of pure functions as linear combinations of MPLs with rational numbers as coefficients.
 
 Pure integrals and MPLs have several nice features. In particular,
if a family of integrals evaluates to a pure combination of MPLs, then it satisfies a particularly simple system of differential equations~\cite{Henn:2013pwa}. It has been known for a long time already that every Feynman integral can be decomposed into a basis of so-called master integrals, and that the master integrals can be computed as the solution to a system of a coupled differential equations~\cite{Kotikov:1990kg,Remiddi:1997ny,Gehrmann:1999as}. The differential equations technique has received new impetus in ref.~\cite{Henn:2013pwa}, where it was conjectured that, in the case where Feynman integrals evaluate to MPLs, it is possible to find a basis of master integrals that are pure integrals. The latter satisfy a system of differential equations that can easily by solved in terms of iterated integrals. This new insight has led to breakthroughs in the analytic computation of multi-loop Feynman integral, cf., e.g., refs.~\cite{Henn:2013fah,Gehrmann:2013cxs,Gehrmann:2014bfa,Henn:2013woa,Henn:2013nsa,DiVita:2014pza,Henn:2014lfa,Caron-Huot:2014lda,Caola:2014lpa,Grozin:2014hna,Bonciani:2015eua,Grozin:2015kna,Gehrmann:2015bfy,Henn:2016men,Henn:2016jdu,Henn:2016kjz,Lee:2016ixa,Bonciani:2016ypc,DiVita:2017xlr,Mastrolia:2017pfy,DiVita:2018nnh,Gehrmann:2018yef,Chicherin:2018mue}.

The concept of pure integrals is not only important to perform explicit computations. Certain QFTs exhibit the feature that some observables can be expressed in terms of pure combinations of MPLs. More precisely, to every MPL one can associate an `invariant' called its \emph{weight}, corresponding to the number of iterated integrations in its definition. It is conjectured that for certain QFTs
like the $\cN=4$ Super Yang-Mills (SYM) theory quantum corrections often evaluate to pure combinations of MPLs of uniform weight. This conjecture is supported by many explicit results in $\cN=4$ SYM not only for scattering amplitudes~\cite{Bern:1994zx,Bern:1994cg,Bern:1997nh,Bern:2004ky,Anastasiou:2003kj,Bern:2004bt,Bern:2005iz,DelDuca:2009au,DelDuca:2010zg,Goncharov:2010jf,Dixon:2011nj,Dixon:2011pw,Dixon:2012yy,Dixon:2013eka,Dixon:2014iba,Dixon:2014voa,Drummond:2014ffa,Dixon:2015iva,Caron-Huot:2016owq,Dixon:2016nkn,DelDuca:2016lad,Henn:2016jdu,DelDuca:2018hrv}, but also for certain anomalous dimensions~\cite{Kotikov:2001sc,Kotikov:2002ab,Kotikov:2004er,Kotikov:2007cy,Beisert:2006ez,Grozin:2015kna,Almelid:2015jia,Dixon:2017nat}, form factors~\cite{vanNeerven:1985ja,Gehrmann:2011xn,Brandhuber:2012vm,Brandhuber:2014ica}, correlation functions~\cite{Eden:1998hh,GonzalezRey:1998tk,Eden:2000mv,Bianchi:2000hn,Drummond:2013nda} and correlators of semi-infinite Wilson lines~\cite{Li:2014bfa,Li:2014afw}. This overwhelming list of results hints towards the fact that the property of uniform weight is not coincidental, but an intrinsic mathematical feature of the theory. Understanding this feature in more detail may not only shed light onto the properties of $\cN=4$ SYM, but on the mathematical structure of multi-loop computations and QFT in general.

It has been known for a long time, however, that not every Feynman integral can be expressed in terms of MPLs. The first time a non-polylogarithmic function was observed in a perturbative result was in the two-loop corrections to the electron self-energy in QED~\cite{Sabry}. Since then, non-polylogarithmic functions were observed to appear also in several other higher-order computations~\cite{Broadhurst:1987ei,Bauberger:1994by,Bauberger:1994hx,Laporta:2004rb,Kniehl:2005bc,Aglietti:2007as,Czakon:2008ii,CaronHuot:2012ab,Nandan:2013ip,Bloch:2013tra,Brown:2013hda,Remiddi:2013joa,Adams:2013nia,Huang:2013kh,Hauenstein:2014mda,Sogaard:2014jla,Adams:2014vja,Adams:2015gva,Adams:2015ydq,Georgoudis:2015hca,Remiddi:2016gno,Bonciani:2016qxi,Bloch:2016izu,Adams:2016xah,Passarino:2016zcd,Broedel:2017siw,vonManteuffel:2017hms,Primo:2017ipr,Ablinger:2017bjx,Chen:2017pyi,Bourjaily:2017bsb,Chen:2017soz,Broedel:2017kkb,Hidding:2017jkk,Becchetti:2017abb,Mistlberger:2018etf,Broedel:2018iwv,Adams:2018bsn,Adams:2018kez,Broedel:2018rwm,Adams:2018ulb,Blumlein:2018aeq,Blumlein:2018jgc,Vanhove:2018mto,Bourjaily:2018ycu}. In many cases these new classes of functions are related to elliptic curves (though it is known that also more complicated geometric structures like Riemann surfaces of higher genus or K3 surfaces appear in QFT~\cite{Brown:2010bw,Huang:2013kh,Hauenstein:2014mda,Bloch:2014qca,Georgoudis:2015hca,Bloch:2016izu,Bourjaily:2018ycu}). For this reason, a lot of effort has recently gone into understanding in more detail the mathematical properties of the classes of functions of elliptic type that show up in multi-loop computations.

From a mathematical point of view, (part of) the family of functions relevant to elliptic Feynman integrals seem to be the so-called \emph{elliptic multiple polylogarithms} (eMPLs)~\cite{BrownLevin}. The eMPLs of ref.~\cite{BrownLevin} are defined as iterated integrals on an elliptic curve defined as a complex torus. Incidentally, the same class of functions is known to describe also scattering amplitudes in superstring theory~\cite{Broedel:2014vla,Broedel:2015hia,Broedel:2017jdo,Broedel:2018izr,Schlotterer:2018zce}, where complex tori and related surfaces naturally occur as the worldsheet relevant to one-loop computations. In ref.~\cite{Broedel:2017kkb} it was shown how eMPLs can equivalently be described as iterated integrals on an elliptic curve defined by a polynomial equation. This latter description is often more convenient when working with elliptic Feynman integrals, because the explicit algebraic description of the elliptic curve is more directly related to the kinematics of the process and Feynman parameter integrals. Closely related to eMPLs are iterated integrals of modular forms~\cite{ManinModular,Brown:mmv}, and it was observed that they also show up in Feynman integral computations~\cite{Ablinger:2017bjx,Adams:2017ejb,Broedel:2018iwv,Adams:2018ulb}. 

Despite all this progress in computing Feynman integrals that do not evaluate to ordinary MPLs, we are still lacking an analogue of some the technology and understanding of MPLs. In particular, it is not entirely clear how to extend the notion of pure functions to the elliptic case. Pure functions have played an important role in modern approaches to Feynman integrals via the differential equations technique described above~\cite{Henn:2013pwa}. It was observed that in some cases the differential equations satisfied by elliptic Feynman integrals can be cast in a form very reminiscent of the non-elliptic cases~\cite{Remiddi:2016gno,Adams:2018yfj}. Since the modern approach to differential equations for Feynman integrals heavily relies on the concept of pure functions, a better understanding of purity in the elliptic case is likely to shed light on this. Moreover, elliptic functions also show up in $\cN=4$ SYM~\cite{CaronHuot:2012ab,Nandan:2013ip,Bourjaily:2017bsb}. Having an understanding of the concept of pure functions would allow one to investigate in how far the conjectured property of uniform weight carries over to elliptic cases.

The purpose of this paper is to take some first steps in trying to understand how to define pure functions of uniform weight in the context of elliptic Feynman integrals. By analysing explicit results for Feynman integrals that can be evaluated in terms of eMPLs~\cite{trianglepaper}, we observe that in all cases the results can be cast in the form of a combination eMPLs of uniform weight. In order to arrive at this result, we need to define a new basis for the space of all eMPLs which is characterised by the fact that all basis elements, seen as functions in many variables, satisfy a differential equation without homogeneous term and have at most logarithmic singularities. These properties are the natural generalisation of the characteristic properties of ordinary MPLs, and we therefore propose that these properties are the natural criteria to demand from a pure function. 

This paper is organized as follows. In Section~\ref{sec:mpls} we give a short review of ordinary MPLs and pure functions that can be expressed in terms of them. We review the background on elliptic curves and eMPLs needed throughout this paper in Section~\ref{sec:empls}. In Section~\ref{sec:pure_empls} we motivate and introduce our class of eMPLs that define pure functions and summarise some of their properties. In Section~\ref{sec:applications} we illustrate these concepts on a selected set of elliptic Feynman integrals. Finally, in Section~\ref{sec:pure_building_blocks} we propose how to define a notion of weight on the functions that appear in elliptic Feynman integral, and in Section~\ref{sec:conclusions} we draw our conclusions. We also include some appendices with some technical material omitted in the main text.


\section{Feynman integrals, pure functions and multiple polylogarithms}
\label{sec:mpls}

We start by reviewing the concept of pure functions in the context of ordinary multiple polylogarithms (MPLs). MPLs are a class of special functions defined by~\cite{Lappo:1927,Goncharov:1998kja,GoncharovMixedTate}
 \beq\label{eq:Mult_PolyLog_def}
 G(a_1,\ldots,a_n;x)=\,\int_0^x\,\frac{d t}{t-a_1}\,G(a_2,\ldots,a_n;t)\,,
\eeq
 and the recursion starts with $G(;x)\equiv 1$. In the special case where all the $a_i$'s are zero, we define
\beq\label{eq:GLog}
G(\underbrace{0,\ldots,0}_{n\textrm{ times}};x) = \frac{1}{n!}\,\ln^n x\,.
\eeq
MPLs satisfy many identities. In particular, they are invariant under a simultaneous rescaling of all arguments,
\beq\label{eq:rescaling}
G(p\,a_1,\ldots,p\,a_n;p\,x) =  G(a_1,\ldots,a_n;x)\,,\quad p, a_n\neq0\,.
\eeq
They form a shuffle algebra
\beq\label{eq:G_shuffle}
G(a_1,\ldots,a_k;x)\,G(a_{k+1},\ldots,a_{k+l};x) = \sum_{\sigma\in \Sigma(k,l)}G(a_{\sigma(1)},\ldots,a_{\sigma(k+l)};x)\,,
\eeq
where $\Sigma(k,l)$ denotes the set of all shuffles of $(a_1,\ldots,a_k)$ and $(a_{k+1},\ldots,a_{k+l})$, i.e., the set of all permutations of their union that preserve the relative orderings within each set. There is a systematic understanding of how to handle such identities (at least in cases relevant to physics applications), see e.g. refs.~\cite{Goncharov:2010jf,Duhr:2011zq,Duhr:2012fh,Duhr:2014woa}.

MPLs are the prototypical examples of a class of functions dubbed \emph{pure functions} in the physics literature. Since one of the main goals of this paper is to extend the notion of pure functions to the elliptic case, we spend some time reviewing this concept in detail. In ref.~\cite{ArkaniHamed:2010gh} a \emph{pure integral} is defined as an integral such that all non-vanishing residues of its integrand are the same up to a sign (in which case we may normalise the integral such that all non-vanishing residues are $\pm1$). A closely related definition uses the notion of {weight}: a \emph{pure function of weight $n$} is a function whose total differential can be written in terms of pure functions of weight $n-1$ (multiplied by algebraic functions with at most single poles)~\cite{Henn:2013pwa}. The recursive definition starts with assigning weight zero to algebraic functions. It is easy to check that sums and products of pure functions are still pure, and the weight of a product of two pure functions is the sum of their weights. The concept of weight is extended from functions to numbers in an obvious way, e.g., the weight of $i\pi = \log(-1)$ is one and the weight of $\zeta_n = -G(\vec{0}_{n-1},1;1)$ is $n$ (and $\vec{0}_{n-1} =(\underbrace{0,\ldots,0}_{n-1\textrm{ times}})$).

MPLs are pure functions with respect to either of these two definitions. Indeed, it is easy to see that all non-vanishing residues of the integrand in eq.~\eqref{eq:Mult_PolyLog_def} are $\pm1$, and so eq.~\eqref{eq:Mult_PolyLog_def} defines a pure integral. Moreover, MPLs satisfy the differential equation~\cite{Goncharov:1998kja}
\beq\bsp\label{eq:MPL_tot_diff}
d G(a_1,\ldots,a_n;z)&\, = \sum_{i=1}^nG(a_1,\ldots,\hat a_i,\ldots,a_n;z)\,d\ln{a_{i-1}-a_i\over a_{i+1}-a_i}\,,
\esp\eeq
from where it immediately follows that $G(a_1,\ldots,a_n;z)$ is a pure function of weight $n$. In the equation above, the notation $\hat{a}_i$ indicates that the label $a_i$ is absent. It is easy to see that the weight of an MPL agrees with the number $n$ of iterated integrations in eq.~\eqref{eq:Mult_PolyLog_def}, and that the shuffle product on MPLs in eq.~\eqref{eq:G_shuffle} preserves the weight. We mention here that, conjecturally, there are no relations among MPLs of different weights. We see that MPLs are the prototypical examples of pure functions. Conversely, it is easy to see that if an integral can be evaluated in terms of algebraic functions and MPLs, then this integral is pure if and only if it can be written as a linear combination of (products of) MPLs whose coefficients are rational \emph{numbers}.

While it is well known that loop integrals can often be expressed in terms of MPLs multiplied by algebraic functions, there is naively no reason to believe that Feynman integrals evaluate to pure functions.
In ref.~\cite{Henn:2013pwa} it was argued that, while indeed Feynman integrals usually do not evaluate to pure functions, one can often make a choice of basis that expresses all members of a given family of Feynman integrals through a set of pure integrals (so-called \emph{master integrals}) with algebraic basis coefficients.

In the case of Feynman integrals that evaluate to MPLs, the pure master integrals are characterised by the fact that they have unit leading singularities~\cite{ArkaniHamed:2010gh,Cachazo:2008vp}. Leading singularities are obtained by computing the maximal codimension residues of a Feynman integral, and as such they are closely related to the maximal cut of the integral. In the case of MPLs the basis of pure master integrals can be reached in an algorithmic way~\cite{Henn:2014qga,Lee:2014ioa,Gituliar:2017vzm,Meyer:2016slj}, and the change of basis only involves algebraic functions.

Let us illustrate this on a simple one-loop example, namely the family of the bubble integral with two massive propagators in $D=2-2\eps$ dimensions,
\beq
B_{n_1n_2}(p^2,m_1^2,m_2^2) = e^{\gamma_E\eps}\,\int\frac{d^Dk}{i\pi^{D/2}}\frac{1}{(k^2-m_1^2)^{n_1}((k+p)^2-m_2^2)^{n_2}}\,,
\eeq
where $\gamma_E = -\Gamma'(1)$ is the Euler-Mascheroni constant. Using integration-by-parts (IBP) identities~\cite{Chetyrkin:1981qh,Tkachov:1981wb}, one can show that every integral in this family can be written as a linear combination of three master integrals, which we may choose as
\beq\bsp\label{eq:one-loop}
B_{10}(p^2,m^2,0) &\,= B_{01}(p^2,0,m^2) = e^{\gamma_E\eps}\,\int\frac{d^Dk}{i\pi^{D/2}}\frac{1}{k^2-m^2} \\
&\,= -\frac{1}{\eps}+\log m^2+\ord(\eps)\,,\\
B_{11}(p^2,m_1^2,m_2^2) &\,= e^{\gamma_E\eps}\,\int\frac{d^Dk}{i\pi^{D/2}}\frac{1}{(k^2-m_1^2)\,((k+p)^2-m_2^2)} \\
&\,= \frac{1}{p^2\,(w-\bar{w})}\,\log\left(\frac{\bar{w}(1-w)}{w(1-\bar{w})}\right)+\ord(\eps)\,,
\esp\eeq
with $w\bar{w}=m_1^2/p^2$ and $(1-w)(1-\bar{w})=m_2^2/p^2$.
We see that the master integrals are not pure, because the logarithms are multiplied by algebraic prefactors. We can however easily define a new set of pure master integrals via the following algebraic change of basis,
\beq\label{eq:bubble_rotation}
\left(\begin{array}{c}{B}_{10} \\ {B}_{01} \\ {B}_{11}\end{array}\right) 
=
\left(\begin{array}{ccc}
-1/\eps & 0 & 0 \\ 0 & -1/\eps & 0 \\ 0& 0& -2/(\eps\,p^2(w-\bar{w}))\end{array}\right)\left(\begin{array}{c}\widetilde{B}_{10} \\ \widetilde{B}_{01} \\ \widetilde{B}_{11}\end{array}\right) \,.
\eeq
The functions $\widetilde{B}_{ij}$ are pure, i.e., the coefficient of $\eps^k$ is a linear combination of terms of uniform weight $k$.
The algebraic prefactors in the matrix in eq.~\eqref{eq:bubble_rotation} can be obtained by analysing the cuts of the integrals. For example, the algebraic factor appearing in the expression for the one-loop bubble integral corresponds to the maximal cut of the integral,
\beq
\textrm{Cut}\left[B_{11|D=2}\right] = -\frac{2}{p^2\,(w-\bar{w})}\,,
\eeq
and so we can write
\beq\label{eq:cut_bubble_MPL}
B_{11} = \textrm{Cut}\left[B_{11|D=2}\right]\times\left[-\frac{1}{2}\log\left(\frac{\bar{w}(1-w)}{w(1-\bar{w})}\right)+\ord(\eps)\right]\,.
\eeq

Having a basis of pure master integrals is not only of formal interest, but it also facilitates their computation. Indeed, pure master integrals are expected to satisfy a system of first order differential equations in \emph{canonical form}~\cite{Henn:2013pwa}. 
In the case of MPLs the concept of purity and uniform weight has changed the way we think about Feynman integrals, and it has led to breakthroughs in the computation of master integrals. The extension of these ideas to Feynman integrals that evaluate to elliptic functions is still largely unclear. One of the goals of this paper is to introduce a class of elliptic polylogarithms that may be called pure, and we argue that many of the properties of pure Feynman integrals carry over to the elliptic case. Before introducing these functions, we give a short review of elliptic curves and elliptic polylogarithms in the next section.


\section{Elliptic curves and elliptic polylogarithms}
\label{sec:empls}
In this section we present the minimal mathematical background on elliptic curves and polylogarithms required in the rest of this paper. In the first subsection we focus on elliptic curves, and review elliptic polylogarithms in the next subsection.

\subsection{Elliptic curves}
\label{sec:elliptic_curves}

For our purposes it is sufficient to think of an elliptic curve as the zero set of a polynomial equation of the form $y^2 = P_n(x)$, where $P_n$ is a polynomial of degree $n=3$ or $4$, i.e., the set of points in $\mathbb{CP}^2$ with homogeneous coordinates $[x,y,1]$ constrained by $y^2 = P_n(x)$. We only discuss here the case of a quartic polynomial, $n=4$, since this case appears most commonly in the context of Feynman integrals. Extending the results of the paper to $n=3$ is straightforward. In the following we always assume that the polynomial defining the elliptic curve is given in the form $P_4(x) = (x-a_1)\cdots (x-a_4)$. The roots $a_i$ of $P_4$ are often referred to as the \emph{branch points} of the elliptic curve. For concreteness, we assume in the following that the branch points are real, distinct and ordered according to $a_1<a_2<a_3<a_4$.

There are certain `invariants' that we can attach to an elliptic curve. The most prominent ones are the two \emph{periods} of the elliptic curve. They are defined by
\beq
\omega_1 = 2\,c_{4}\int_{a_2}^{a_3}\frac{dx}{y} = 2\,\textrm{K}(\lambda) 
\textrm{~~~~~~and~~~~~~}\omega_2 = 2\,c_{4}\int_{a_1}^{a_2}\frac{dx}{y} = 2i\,\textrm{K}(1-\lambda)\,,
\label{eq:periods4}
\eeq
with 
\beq\label{eq:lambda4}
\lambda = \frac{a_{14}\,a_{23}}{a_{13}\,a_{24}} \qquad \textrm{and}\qquad
c_{4} = \frac{1}{2}\sqrt{a_{13}a_{24}}\,,\qquad a_{ij} = a_i-a_j\,,
\eeq
and K denotes the complete elliptic integral of the first kind,
\beq
\textrm{K}(\lambda) = \int_0^1\frac{dt}{\sqrt{(1-t^2)(1-\lambda t^2)}}\,.
\eeq 
In eq.~\eqref{eq:periods4} we use the following convention for the branches of the square root in the integrand (valid when the branch points are real), 
\beq\bsp
\sqrt{P_4(x)}&\,\equiv
\sqrt{|P_4(x)|}\times\left\{\begin{array}{ll}
-1\,,& x\le a_1\textrm{ or }x > a_4\,,\\
-i\,,& a_1<x\le a_2\,,\\
\phantom{-}1\,,& a_2<x\le a_3\,,\\
\phantom{-}i\,,& a_3<x\le a_4\,.
\end{array}\right.
\esp \label{eq:rsigns}
\eeq
There are also two \emph{quasi-periods} attached to every elliptic curve. We choose them as
\beq\bsp\label{eq:quasi-periods4}
\eta_1 &\,= -\frac{1}{2}\int_{a_2}^{a_3}dx\,\widetilde{\Phi}_4(x,\vec a) = \textrm{E}(\lambda) -\frac{2-\lambda}{3}\,\textrm{K}(\lambda)\,,\\
\eta_2 &\,= -\frac{1}{2}\int_{a_1}^{a_2}dx\,\widetilde{\Phi}_4(x,\vec a)= -i\,\textrm{E}(1-\lambda) +i\,\frac{1+\lambda}{3}\,\textrm{K}(1-\lambda)\,,
\esp\eeq
where E denotes the complete elliptic integral of the second kind,
\beq
\textrm{E}(\lambda) = \int_0^1dt\,\sqrt{\frac{1-\lambda t^2}{1-t^2}}\,,
\eeq
and $\widetilde{\Phi}_4(x,\vec a)$ is defined by
\beq\label{eq:tilde_Phi_4_def}
\widetilde{\Phi}_4(x,\vec a) \equiv \frac{1}{c_4\,y} \left( x^2 - \frac{s_1}{2}\,x + \frac{s_2}{6} \right)\,.
\eeq
Here $\vec a\equiv(a_1,a_2,a_3,a_4)$ and $s_n \equiv s_n(\vec a)$ denotes the symmetric polynomial of degree $n$ in the branch points. 
The periods and quasi-periods are not independent, but they are related by the Legendre relation,
\beq\label{eq:Legendre}
\eta_2\,\omega_1-\eta_1\omega_2 = -i\pi\,.
\eeq
The function $\widetilde{\Phi}_4$ has the property that the differential one-form $dx\,\widetilde{\Phi}_4(x,\vec a)$ has a double-pole with vanishing residue at $x=\infty$. 

The periods and quasi-periods are not genuine invariants of an elliptic curve, in the sense that isomorphic elliptic curves may give rise to different (quasi-)periods. A true invariant that uniquely characterises every elliptic curve is the so-called $j$-invariant. Since the $j$-invariant will not play any role in this paper, we do not define it here and refer to the literature. 
Instead, we note that the redundancy in the definition of the periods and quasi-periods is due to the fact that they only depend on the cross-ratio $\lambda$ of the four branch points.
Hence, different polynomials may describe the same elliptic curve. In order to resolve this redundancy it is convenient to look at an alternative way to classify elliptic curves.

It can be shown that every elliptic curve defined over the complex numbers (which means that we are looking for complex solutions to the polynomial equation $y^2 = P_4(x)$) is isomorphic to a complex torus, i.e., the quotient of the complex plane $\mathbb{C}$ by a two-dimensional lattice $\Lambda$. In our case the relevant lattice is the lattice $\Lambda=\omega_1\mathbb{Z} + \omega_2\mathbb{Z}$ spanned by the two periods. We can perform a rescaling without changing the geometry, and from now on we will always be working with the torus defined by the lattice $\Lambda_\tau = \mathbb{Z}+\tau\mathbb{Z}$, where $\tau=\omega_2/\omega_1$ denotes the ratio of the two periods, with $\textrm{Im }\tau>0$. In other words, every $\tau$ in the upper half-plane $\mathbb{H} = \{\tau\in\mathbb{C}:\textrm{Im }\tau>0\}$ defines a two-dimensional lattice, and thus an elliptic curve. Different values of $\tau$ may still define the same elliptic curve. One can show that $\tau,\tau'\in\mathbb{H}$ define the same elliptic curve if and only if they are related by a \emph{modular transformation}, i.e., a M\"obius transformation for $SL(2,\mathbb{Z})$. The space of geometrically-distinct tori (the so-called \emph{moduli space}) can then be identified with the quotient of the upper half-plane $\mathbb{H}$ by the modular group $SL(2,\mathbb{Z})$.

The map from the torus $\mathbb{C}/\Lambda_{\tau}$ to the curve defined by the polynomial equation $y^2=P_4(x)$ can be explicitly realised. One can show that there is a function $\kappa(.,\vec a):\mathbb{C}/\Lambda_{\tau}\to \mathbb{C}$ which satisfies the differential equation $(c_4\kappa'(z,\vec a))^2=P_4(\kappa(z,\vec a))$, and the image of the torus under $\kappa$ can be identified with the elliptic curve $y^2=P_4(x)$. The explicit form of $\kappa$ is not important in the following, and we refer to ref.~\cite{Broedel:2017kkb} for its explicit definition. Here we only mention that $\kappa$ is a meromorphic function of $z$ and it is doubly-periodic, that is $\kappa(z+1,\vec a)=\kappa(z+\tau,\vec a)=\kappa(z,\vec a)$. A function satisfying these properties is called an \emph{elliptic function}. Moreover, $\kappa$ is an even function of $z$, and it maps the half-periods to the branch points $a_i$,
\beq\label{eq:kappa_half_periods}
\kappa(0,\vec a) = a_1\,,\quad \kappa(\tau/2,\vec a) = a_2\,,\quad \kappa(1/2+\tau/2,\vec a) = a_3\,,\quad \kappa(1/2,\vec a) = a_4\,.
\eeq
The inverse map to $\kappa$ is called Abel's map and is defined in the following way. If $[X,Y,1]\in\mathbb{CP}^2$ is a point satisfying $Y^2=P_4(X)$, then its image on the torus is 
\beq\label{eq:Abel}
z_X = \frac{c_4}{\omega_1}\int_{a_1}^X\frac{dx}{y} = \frac{\sqrt{a_{13}a_{24}}}{4\,\textrm{K}(\lambda)}\int_{a_1}^X\frac{dx}{y}\,.
\eeq
In the following an important role will be played by the image $z_{\ast}$ on the torus of the point $x=-\infty$, defined by
\beq\label{eq:zstar_int}
z_{\ast} = \frac{c_4}{\omega_1}\int_{a_1}^{-\infty}\frac{dx}{y} = \frac{\sqrt{a_{13}a_{24}}}{4\,\textrm{K}(\lambda)}\int_{a_1}^{-\infty}\frac{dx}{y}\,.
\eeq
In the case where the branch points $a_1<a_2<a_3<a_4$ are all real and the branches of the square root are chosen according to eq.~\eqref{eq:rsigns}, we can evaluate the integral in terms of elliptic integrals of the first kind (see Appendix~\ref{app:zstar}) to obtain a closed analytic expression for $z_{\ast}\,$. We find
\beq\label{eq:zstar}
z_{\ast} = \cZ_{\ast}(\alpha,\lambda)\equiv  \frac{1}{2} - \frac{\textrm{F}(\sqrt{\alpha} | \lambda)}{ 2\, \textrm{K}(\lambda)}\,, \qquad \alpha = \frac{a_{13}}{a_{14}}\,,
\eeq
where F denotes the incomplete elliptic integral of the first kind,
\beq\label{eq:F_def}
\textrm{F}(x|\lambda) = \int_0^x\frac{dt}{\sqrt{(1-t^2)(1-\lambda t^2)}}\,.
\eeq 
In other situations, e.g., when the branch points $a_i$ are not real and/or the branches of the square root are not chosen according to eq.~\eqref{eq:rsigns}, then eq.~\eqref{eq:zstar} still remains true and we can express $z_{\ast}$ in terms of $\cZ_{\ast}(\alpha,\lambda)$, albeit only up to complex conjugation and to a sign related to the choice of the branches of the square root. We refer to Appendix~\ref{app:zstar} for a detailed discussion.

\subsection{Elliptic multiple polylogarithms}
\label{subsec:empls}
In this subsection we introduce a generalisation of polylogarithms to elliptic curves. We start by defining elliptic multiple polylogarithms as iterated integrals on (the universal cover of) a complex torus, and we review what this class of integrals becomes in terms of the variables $(x,y)$ at the end of this subsection.

Elliptic multiple polylogarithms (eMPLs) were first introduced in ref.~\cite{BrownLevin}. Here we follow a slightly different path, and inspired by refs.~\cite{MatthesThesis,Broedel:2014vla} we define eMPLs by the iterated integral
\beq\label{eq:gamt_def}
\gamtt{n_1 &\ldots& n_k}{z_1 & \ldots & z_k}{z}{\tau} = \int_0^zdz'\,g^{(n_1)}(z'-z_1,\tau)\,\gamtt{n_2 &\ldots& n_k}{z_2 & \ldots & z_k}{z'}{\tau}\,,
\eeq
where $z_i$ are complex numbers and $n_i\in \mathbb{N}$ are positive integers. 
The integers $k$ and $\sum_in_i$ are called the \emph{length} and the \emph{weight} of the eMPL. In the case where $(n_k,z_k) = (1,0)$, the integral in eq.~\eqref{eq:gamt_def} is divergent and requires regularisation. We closely follow ref.~\cite{Broedel:2014vla} for the choice of the regularisation scheme. 

The integration kernels in eq.~\eqref{eq:gamt_def} are defined through a generating series known as the \emph{Eisenstein-Kronecker series},
\beq\label{eq:Eisenstein-Kronecker}
F(z,\alpha,\tau) = \frac{1}{\alpha}\,\sum_{n\ge0}g^{(n)}(z,\tau)\,\alpha^n = \frac{\theta'_1(0,\tau)\,\theta_1(z+\alpha,\tau)}{\theta_1(z,\tau)\,\theta_1(\alpha,\tau)}\,,
\eeq
where $\theta_1$ is the odd Jacobi theta function, and $\theta'_1$ is its derivative with respect to its first argument. Seen as a function of $z$, the function $g^{(1)}(z,\tau)$ has a simple pole with unit residue at every point of the lattice $\Lambda_{\tau}$. For $n>1$, $g^{(n)}(z,\tau)$ has a simple pole only at those lattice points that do not lie on the real axis. As a consequence, the iterated integrals in eq.~\eqref{eq:gamt_def} have at most logarithmic singularities. 
Furthermore the functions $g^{(n)}$ have definite parity,
\beq\label{eq:g_parity}
g^{(n)}(-z,\tau) = (-1)^n\,g^{(n)}(z,\tau)\,,
\eeq
and are invariant under translations by 1, but not $\tau$,
\beq\label{eq:g_periodic}
g^{(n)}(z+1,\tau) = g^{(n)}(z,\tau) \textrm{~~and~~}  g^{(n)}(z+\tau,\tau) = \sum_{k=0}^n\frac{(-2\pi i)^k}{k!}\,g^{(n-k)}(z,\tau)\,.
\eeq

Elliptic MPLs share many of the properties of ordinary MPLs. First, eMPLs form a shuffle algebra, 
\beq
\widetilde{\Gamma}(A_1 \cdots A_k;z,\tau)\,\widetilde{\Gamma}(A_{k+1}\cdots A_{k+l};z,\tau) = \sum_{\sigma\in \Sigma(k,l)}\widetilde{\Gamma}(A_{\sigma(1)} \cdots A_{\sigma(k+l)};z,\tau)\,,
\eeq
where we have introduced the notation $A_i=\left(\begin{smallmatrix}n_i\\ z_i\end{smallmatrix}\right)$. The shuffle product preserves both the weight and the length of eMPLs. Second, there is a closed formula for the total differential of an eMPL which is very reminiscent of the total differential of an ordinary MPL in eq.~\eqref{eq:MPL_tot_diff}. The formula for the total differential reads~\cite{Broedel:2018iwv},
\beq\bsp\label{eq:gamma_differential}
d\widetilde{\Gamma}&\left(A_1\cdots A_k;z,\tau \right) = \sum_{p=1}^{k-1}(-1)^{n_{p+1}}\,\widetilde{\Gamma}\left(A_1\cdots A_{p-1}\; ^0 _0 \; A_{p+2}\cdots A_k;z,\tau \right)\,\omega_{p,p+1}^{(n_p+n_{p+1})}\\
&\,+\sum_{p=1}^{k}\sum_{r=0}^{n_p+1}\Bigg[\binom{n_{p-1}+r-1}{n_{p-1}-1}\,\widetilde{\Gamma}\left(A_1\cdots A_{p-1}^{[r]} \; \hat{A}_{p} \; A_{p+1}\cdots A_k;z,\tau \right)\,\omega_{p,p-1}^{(n_p-r)}\\
&\,\phantom{\sum_{p=1}^{k}\sum_{r=0}^{n_p+1}\Big[}
-\binom{n_{p+1}+r-1}{n_{p+1}-1}\,\widetilde{\Gamma}\left(A_1\cdots A_{p-1} \; \hat{A}_{p} \; A_{p+1}^{[r]}\cdots A_k;z,\tau \right)\,\omega_{p,p+1}^{(n_p-r)}\Bigg]\,,
\esp\eeq
where, similarly to the case of MPLs, the hat indicates that the corresponding argument is absent and we have introduced the shorthands
\beq\bsp 
A_i^{[r]}\,\equiv\, \left( \begin{smallmatrix}
	n_i+r \\ z_i
\end{smallmatrix}\right) {\rm~~and~~} A_i^{[0]}\,\equiv\,A_i\,.
\esp\eeq
In the previous equation, we let $(z_0,z_{k+1})=(z,0)$ and $(n_0,n_{k+1})=(0,0)$, and we use the convention that the binomial number $\binom{-1}{-1}$ is $1$. The differential one-forms $\omega_{ij}^{(n)}$ are given by
\beq\bsp\label{eq:empl_letter}
\omega_{ij}^{(n)} &\,\equiv \omega^{(n)}(z_j-z_i)= (dz_j-dz_i)\,g^{(n)}(z_j-z_i,\tau) + \frac{n\,d\tau}{2\pi i}\,g^{(n+1)}(z_j-z_i,\tau)\,,
\esp\eeq
with $g^{(-1)}(z,\tau) = 0$. 
We note here that both ordinary and elliptic MPLs satisfy a differential equation without homogeneous term, as can be seen from eqs.~\eqref{eq:MPL_tot_diff} and \eqref{eq:gamma_differential}. Functions of this type are called \emph{unipotent}. The differential equation satisfied by unipotent functions serves as the basis to define a symbol map and a coaction on them~\cite{Brown:coaction}. The coaction decomposes every MPL (elliptic or not) into a tensor product whose first entry is itself an MPL, while the second entry is interpreted as a symbol of sorts. In the non-elliptic case, this coaction is closely related to the coaction on ordinary MPLs~\cite{Goncharov:2005sla,GoncharovMixedTate,Brown:2011ik} (see also ref.~\cite{Duhr:2012fh,Duhr:2014woa}). Details about this construction in the case of eMPLs can be found in ref.~\cite{Broedel:2018iwv}.

Not all the functions encountered when working with elliptic curves are unipotent. In particular, the periods and quasi-periods in eq.~\eqref{eq:periods4} and~\eqref{eq:quasi-periods4} give rise to non-unipotent functions. To see how they arise, it is convenient to combine the periods and quasi-periods into a $2\times2$ matrix
\beq\label{eq:per_mat}
P=\left(\begin{array}{cc}
\omega_1&\omega_2\\
\eta_1&\eta_2
\end{array}\right)\,,
\eeq
The Legendre relation in eq.~\eqref{eq:Legendre} reduces to $\det P= -i\pi$.
We can write this matrix as the product of two matrices, $P=SU$, with
\beq\label{eq:US_matrix}
S=\left(\begin{array}{cc}
\omega_1&0\\
\eta_1&-i\pi/\omega_1
\end{array}\right)
\textrm{~~and~~}
U=\left(\begin{array}{cc}
1&\tau\\
0&1
\end{array}\right)\,.
\eeq
We stress that this factorisation is based on a choice, because we have singled out $\omega_1$ with respect to $\omega_2$. While from a purely mathematically standpoint there is no natural way to prefer $\omega_1$ over $\omega_2$, we can often appeal to physics to motivate the choice (e.g., because it is often possible to choose $\omega_1$ to be real and $\omega_2$ to be purely imaginary, at least in some region of kinematic space).
It is easy to check that the matrix $U$ satisfies a differential equation without homogeneous term, and so $\tau$ is unipotent. The elements of $S$ instead are not unipotent and referred to as \emph{semi-simple}, i.e., in our case the quantities $i\pi/\omega_1$, $\omega_1$ and $\eta_1$ are not unipotent, but semi-simple. 

So far we have described eMPLs as iterated integrals on a complex torus. Since we can map the torus to the elliptic curve defined by the polynomial equation $y^2=P_4(x)$, we can obtain an alternative description of eMPLs as iterated integrals directly in the coordinates $(x,y)$. This was worked out explicitly for the first time in ref.~\cite{Broedel:2017kkb}, where the following class of iterated integrals was defined,
\beq\label{eq:E4_def}
\Efe{n_1 & \ldots & n_k}{c_1 & \ldots& c_k}{x}{\vec{a}} = \int_0^xdt\,\psi_{n_1}(c_1,t,\vec a)\,\Efe{n_2 & \ldots & n_k}{c_2 & \ldots& c_k}{t}{\vec a}\,,
\eeq
with $n_i\in\mathbb{Z}$ and $c_i\in\widehat{\mathbb{C}}=\mathbb{C}\cup\{\infty\}$, and the recursion starts with $\textrm{E}_4(;x,\vec a)=1$. The integration kernels $\psi_n$ are obtained by explicitly constructing a basis of integration kernels with at most simple poles in $x$ on an elliptic curve. We refer to ref.~\cite{Broedel:2017kkb} for the details, and we content ourselves here to present the subset corresponding to $|n|\le 2$, which (empirically) is the one most relevant to the computation of two-loop Feynman integrals evaluating to eMPLs. The simplest integration kernel $\psi_0$ defines the holomorphic differential on the elliptic curve,
\beq
\psi_0(0,x,\vec a) = \frac{c_4}{y}\,.
\eeq
In particular, the one-form $dx\,\psi_0$ has no pole anywhere on the elliptic curve.
For $n=\pm1$ instead, the kernels have a simple pole at the location specified by the first argument in $\psi_{\pm1}$. They are given by
\beq\bsp\label{eq:psi_1}
\psi_1(c,x,\vec a) = \frac{1}{x-c}\,,\qquad &\psi_{-1}(c,x,\vec a) = \frac{y_c}{y(x-c)}\,, \quad c\neq\infty\,,\\
 \psi_1(\infty,x,\vec a) = \frac{c_4}{y}\,Z_4(x,\vec a)\,,\qquad &\psi_{-1}(\infty,x,\vec a) = \frac{x}{y}\,,
\esp\eeq
where we introduced the shorthand $y_c = \sqrt{P_4(c)}$.
The kernel $\psi_1(c,x,\vec a)$ is identical to the kernel that defines ordinary MPLs, and so ordinary MPLs are a subset of eMPLs.
In eq.~\eqref{eq:psi_1} the function $Z_4$ is defined by the integral
\beq
Z_4(x,\vec a) \equiv \int_{a_1}^xdx'\,\Phi_4(x',\vec a) \,,\qquad \textrm{with} \qquad \Phi_4(x,\vec a)\equiv\widetilde\Phi_4(x,\vec a) +4c_{4}\, \frac{\eta_1}{\omega_1}\,\frac{1}{y}\,,
\eeq
and $\widetilde\Phi_4$ was defined in eq.~\eqref{eq:tilde_Phi_4_def}. Since $dx\,\widetilde\Phi_4$ has a double pole without residue at infinity, the function $Z_4$ has a simple pole there. Similarly, for $|n|>1$ the kernels $\psi_n(c,x,\vec a)$ have at most a simple pole at $x=c$. In addition, they involve higher powers of $Z_4(x,\vec a)$ (while still only having at most simple poles at infinity). For example, we have~\cite{Broedel:2017kkb}
\beq\bsp
\qquad\psi_{-2}(c,x,\vec a)  = \frac{y_c}{y(x-c)}\,Z_4(x,\vec a)\,,\qquad &\psi_2(c,x,\vec a) = \frac{1}{x-c}\,Z_4(x,\vec a)-\Phi_4(x,\vec a)\,, \quad c\neq\infty\,,\\
 \psi_2(\infty,x,\vec a) = \frac{c_4}{y}\,Z_4^{(2)}(x,\vec a)\,,\qquad & \psi_{-2}(\infty,x,\vec a) = \frac{x}{y}\,Z_4(x,\vec a)-\frac{1}{c_4}\,,
\esp\eeq
where $Z_4^{(2)}(x,\vec a)$ is a polynomial of degree two in $Z_4(x,\vec a)$. The concrete form of this polynomial can be found in ref.~\cite{Broedel:2017kkb} and we do not repeat it here because it is irrelevant for the discussion that follows.

We have now two different descriptions of eMPLs, either in terms of the functions $\widetilde{\Gamma}$ or the functions $\textrm{E}_4$. These two classes of functions are in fact just two different bases for the same space of functions. Indeed, it was shown in ref.~\cite{Broedel:2017kkb} that one can always write the kernels $\psi_n$ as linear combinations of the coefficients $g^{(n)}$ of the Eisenstein-Kronecker series. For example, the holomorphic differential can be written as\footnote{The additional factor of $\omega_1$ compared to ref.~\cite{Broedel:2017kkb} comes from the fact that here we work with the torus defined by the lattice $\mathbb{Z}+\tau\mathbb{Z}$ instead of $\omega_1\mathbb{Z}+\omega_2\mathbb{Z}$.} $dx\,\psi_0(0,x,\vec a) = \omega_1\,dz$, where $z$ denotes the image of $x$ under Abel's map in eq.~\eqref{eq:Abel}. The kernels in eq.~\eqref{eq:psi_1} can be related to the kernels defined on the torus as~\cite{Broedel:2017kkb}
\begin{align}
\nonumber dx\,\psi_1(c,x,\vec a) &\,= dz\,\left[g^{(1)}(z-z_c,\tau) + g^{(1)}(z+z_c,\tau) - g^{(1)}(z-z_\ast,\tau) - g^{(1)}(z+z_\ast,\tau)\right]\,,\\
\nonumber dx\,\psi_{-1}(c,x,\vec a) &\,= dz\,\left[g^{(1)}(z-z_c,\tau) - g^{(1)}(z+z_c,\tau) + g^{(1)}(z_c-z_\ast,\tau) + g^{(1)}(z_c+z_\ast,\tau)\right]\,,\\
\label{eq:psi_to_g} dx\,\psi_1(\infty,x,\vec a) &\,= dz\,\left[ - g^{(1)}(z-z_\ast,\tau) - g^{(1)}(z+z_\ast,\tau)\right]\,,\\
\nonumber dx\,\psi_{-1}(\infty,x,\vec a) &\,= \frac{a_1\,\omega_1\,dz}{c_4}+dz\,\left[ g^{(1)}(z-z_\ast,\tau) - g^{(1)}(z+z_\ast,\tau) + 2g^{(1)}(z_\ast,\tau)\right]\,,
\end{align}
where $z_{\ast}$ is defined in eq.~\eqref{eq:zstar}.
Similar formulas can be derived for all other kernels. We refer to ref.~\cite{Broedel:2017kkb} for the details. Using these relations, one can easily check that there is a one-to-one map between the functions $\widetilde{\Gamma}$ and the functions $\textrm{E}_4$, and we can always express a function from one class as a linear combination of functions from the other class. Finally, we note that we can write the function $Z_4$ in terms of the coefficients of the Eisenstein-Kronecker series~\cite{Broedel:2017kkb},
\beq\label{eq:Z4_to_g1}
Z_4(x,\vec a) = -\frac{1}{\omega_1}\left[g^{(1)}(z_x-z_{\ast},\tau) + g^{(1)}(z_x+z_{\ast},\tau)\right]\,.
\eeq


\section{Elliptic polylogarithms and pure functions}
\label{sec:pure_empls}

In the previous sections we have reviewed multiple polylogarithms, both ordinary and elliptic. We have seen that the elliptic and non-elliptic cases share many properties. In this section we argue how the concept of pure functions can be extended from ordinary to elliptic MPLs. Before we present the definition of pure eMPLs, we discuss in the next subsection the motivation for that definition. 

\subsection{Motivation}
A priori, it is not entirely clear how to extend the definition of pure functions to the elliptic case, thus we approach the issue by analysing available results for Feynman integrals that evaluate to eMPLs. 
A naive definition of a pure elliptic Feynman integral could consist in considering $\mathbb{Q}$-linear combinations of elliptic polylogarithms $\textrm{E}_4$ of the same length or weight. Such a naive definition, however, soon reaches its limits, as we now demonstrate. 

In ref.~\cite{Broedel:2017siw} the two-loop sunrise integral in $D=2-2\eps$ with three equal masses was computed in terms of $\textrm{E}_4$ functions. More precisely, consider the family of integrals
\beq
S_{n_1n_2n_3}(p^2,m^2) = -\frac{e^{2\gamma_E\eps}}{\pi^D}\int\frac{d^Dk\,d^Dl}{(k^2-m^2)^{n_1}(l^2-m^2)^{n_2}((k+l+p)^2-m^2)^{n_3}}\,,
\eeq
with $n_i\in \mathbb{N}$.
Using IBP identities, every integral in this family can be written as a linear combination of the following three master integrals,
\beq\bsp\label{eq:sunrise_masters}
S_0(p^2,m^2)&\, = S_{110}(p^2,m^2)\,,\\
 S_1(p^2,m^2) &\,= S_{111}(p^2,m^2)\,,\\
  S_2(p^2,m^2)&\, = S_{112}(p^2,m^2)\,.
\esp\eeq
$S_0$ is the product of two one-loop tadpole integrals and will not be discussed any further. For now, we focus only on the master integral $S_1$, and we return to $S_2$ in Section~\ref{sec:applications}.
The result for $S_1$ reads~\cite{Broedel:2017siw}
\beq\bsp
\label{eqn:S111_result}
S_{1}(p^2,m^2) 
&\,=\frac{1}{(m^2+p^2)c_4}\Biggr[\frac{1}{c_4}\Efe{0&0}{0&0}{1}{\vec a}-2\Efe{0&-1}{0&\infty}{1}{\vec a}-\Efe{0&-1}{0&0}{1}{\vec a}\\
&\,\phantom{=\frac{1}{(m^2+p^2)c_4}\Biggr[}-\Efe{0&-1}{0&1}{1}{\vec a}-\Efe{0&1}{0&0}{1}{\vec a}\Biggr] + \ord(\eps)\,,
\esp\eeq
where the vector of branch points $\vec a$ is
\begin{equation}\begin{split}
  \label{eqn:roots}
\vec a = \left(\frac{1}{2}(1+\sqrt{1+\rho}),\frac{1}{2}(1+\sqrt{1+\overline{\rho}}),\frac{1}{2}(1-\sqrt{1+\rho}),\frac{1}{2}(1-\sqrt{1+\overline{\rho}})\right)\,.
\end{split}\end{equation}
with
\begin{equation}
    \rho = -\frac{4m^2}{(m+\sqrt{-p^2})^2} \textrm{~~and~~} \overline{\rho} = -\frac{4m^2}{(m-\sqrt{-p^2})^2}\,. 
\end{equation}
The result for $S_{1}$ in eq.~\eqref{eqn:S111_result} is not pure (not even up to an overall algebraic factor), because not all the $\textrm{E}_4$ functions are multiplied by rational \emph{numbers}, but the first term in square brackets is multiplied by the algebraic function $1/c_4$. There is, however, strong motivation to believe that the two-loop sunrise integral in $D=2-2\eps$ dimensions should define a pure function of some sort: First, while eq.~\eqref{eqn:S111_result} was obtained by integrating the Feynman parameter representation for $S_1$, the corresponding result obtained from dispersion relations can be written as a $\mathbb{Q}$-linear combination of $\textrm{E}_4$ functions, and no additional algebraic prefactor is needed~\cite{Broedel:2017siw}. Second, in the case where at least one propagator is massless, the integral can be evaluated in terms of pure linear combination of ordinary MPLs. Third, the equal-mass sunrise integral $S_{1}(p^2,m^2)$ can also be written in terms of iterated integrals of Eisenstein series~\cite{Adams:2017ejb,Broedel:2018iwv}, and also in this representation no additional algebraic prefactor is needed. 

Based on this example, we see that a naive definition of `elliptic purity' via the basis of eMPLs $\textrm{E}_4$ does not have the desired properties. Instead of working with the basis of $\textrm{E}_4$ functions, we can change basis and consider the basis of eMPLs $\widetilde{\Gamma}$ on the complex torus. We observe that when expressed in terms of this basis, the equal-mass sunrise integral is a $\mathbb{Q}$-linear combination of $\widetilde{\Gamma}$ functions (up to an overall factor), i.e., it is a function that can be qualified as pure. The final expression for $S_{1}$ in terms of pure functions will be given in Section~\ref{sec:applications}. We therefore propose that the functions $\widetilde{\Gamma}$ are pure functions, and an elliptic Feynman integral is pure if it can be expressed as a $\mathbb{Q}$-linear combination of such functions (up to overall normalisation). As a motivation for this proposal we point out that, just like ordinary MPLs, the functions $\widetilde{\Gamma}$ have at most logarithmic singularities in all variables. Indeed, we have seen in Section~\ref{sec:empls} that the integration kernels $g^{(n)}(z,\tau)$ have at most simple poles, and so the one-forms $\omega_{ij}^{(n)}$ that appear in the total differential of $\gamtt{n_1 &\ldots& n_k}{z_1 & \ldots & z_k}{z}{\tau}$ have at most logarithmic singularities (see eqs.~\eqref{eq:gamma_differential} and~\eqref{eq:empl_letter}). Hence, seen as a function in many variables, $\gamtt{n_1 &\ldots& n_k}{z_1 & \ldots & z_k}{z}{\tau}$ has only logarithmic singularities in each variable, but no poles (because the differential of a pole is a pole of order at least two). This property is identical to the corresponding property for ordinary MPLs, as can easily be seen from the fact that the total differential of an ordinary MPL in eq.~\eqref{eq:MPL_tot_diff} only involves logarithmic one-forms. The $\textrm{E}_4$ functions, instead, do not only have logarithmic singularities when seen as a function of many variables, but also poles. Indeed, consider the function $\Efe{-1}{c}{x}{\vec a}$, with $c\neq\infty$. Using eqs.~\eqref{eq:psi_to_g} and~\eqref{eq:Z4_to_g1} (and assuming for simplicity that $z_0=0$), we find
\begin{align}
\nonumber\Efe{-1}{c}{x}{\vec a} &\,= \gamtt{1}{z_c}{z_x}{\tau} - \gamtt{1}{-z_c}{z_x}{\tau} + \left[g^{(1)}(z_c-z_{\ast},\tau) + g^{(1)}(z_c+z_{\ast},\tau)\right]\,\gamtt{0}{0}{z_x}{\tau}\\
&\,=\gamtt{1}{z_c}{z_x}{\tau} - \gamtt{1}{-z_c}{z_x}{\tau} -\omega_1\,Z_4(c,\vec a) \,\gamtt{0}{0}{z_x}{\tau}\,.
\end{align}
While the functions $\widetilde{\Gamma}$ have at most logarithmic singularities, the function $Z_4(c,\vec a)$ has a pole at $c=\infty$. Hence, unlike the $\widetilde{\Gamma}$ functions, the $\textrm{E}_4$ functions have poles. We stress, however, that in the variable $x$ they only have logarithmic singularities, because the kernels $\psi_n(c,x,\vec a)$ have at most simple poles in $x$. Based on these considerations we propose the following definition of {pure functions}:
\begin{quote}
\emph{A function is called pure if it is unipotent and its total differential involves only pure functions and one-forms with at most logarithmic singularities.}
\end{quote}
Sums and product of pure functions are obviously pure. We postulate that a pure function remains pure under (reasonable) specialisations of the arguments to algebraic numbers. This allows us in particular to extend the definition from functions to numbers.

While the functions $\widetilde{\Gamma}$ provide a basis of pure eMPLs, this basis is often not the most convenient one when working with Feynman integrals:
\begin{enumerate}
\item Feynman integrals often have an intrinsic notion of `parity', defined in the following way: Although the final analytic result for a Feynman integral may involve square roots, the original integrand is a purely rational object. Hence, the final analytic result including square roots must be independent of the choice of the branch of the root. This implies that the pure function part must have definite `parity' with respect to the operation of changing the sign of the root.  For example, we see that the one-loop bubble integral in eq.~\eqref{eq:one-loop} is independent of the sign of the square root, and both the algebraic prefactor and the pure function part are odd functions. In the case of eMPLs, changing the sign of the square root corresponds to the operation $(x,y) \leftrightarrow (x,-y)$. Since $(x,y) = (\kappa(z,\vec a),c_4\,\kappa'(z,\vec a))$, this operation corresponds on the torus to changing the sign of $z$. We would thus like to have a basis of pure functions that have definite parity under this operation. The basis $\widetilde{\Gamma}$ does not have this property, and we prefer to work with an alternative basis that makes this symmetry manifest.
\item From the mathematical point of view, elliptic curves and the functions associated to them are most naturally studied in terms of complex tori and the coordinate $z$. Feynman integrals, however, are more naturally expressed in terms of the variables $(x,y)$, because these variables are more directly related to the kinematics of the process under consideration. We would therefore like to have a basis of pure eMPLs formulated directly in terms of the variables $(x,y)$.
\end{enumerate}


\subsection{Pure elliptic multiple polylogarithms}
\label{sec:pure_eMPLs}
In this section we introduce a new class of iterated integrals on the elliptic curve defined by the polynomial equation $y^2=P_4(x)$ with the following properties:
\begin{enumerate}
\item They form a basis for the space of all eMPLs.
\item They are pure.
\item They have definite parity.
\item They manifestly contain ordinary MPLs.
\end{enumerate}
The definition reads
\beq\label{eq:cE4_def}
\cEf{n_1 & \ldots & n_k}{c_1 & \ldots& c_k}{x}{\vec{a}} = \int_0^xdt\,\Psi_{n_1}(c_1,t,\vec a)\,\cEf{n_2 & \ldots & n_k}{c_2 & \ldots& c_k}{t}{\vec a}\,,
\eeq
with $n_i\in\mathbb{Z}$ and $c_i\in\widehat{\mathbb{C}}$. Equation~\eqref{eq:cE4_def} is of course equivalent to the differential equation,
\beq\label{eq:cE4_derivative}
\partial_x\cEf{n_1 & \ldots & n_k}{c_1 & \ldots& c_k}{x}{\vec{a}} = \Psi_{n_1}(c_1,x,\vec{a})\,\cEf{n_2 & \ldots & n_k}{c_2 & \ldots& c_k}{x}{\vec{a}}\,.
\eeq
The length and the weight are specified in analogy with the case of the $\textrm{E}_4$ functions in eq.~\eqref{eq:E4_def}. The integration kernels are defined implicitly through the identity (for $n\ge 0$)
\begin{align}\label{eq:Psi_to_gh}
dx\,&\Psi_{\pm n}(c,x,\vec a)\\
\nonumber& = dz_x\,\left[g^{(n)}(z_x-z_c,\tau)  \pm g^{(n)}(z_x+z_c,\tau) - \delta_{\pm n,1}\,\left(g^{(1)}(z_x-z_\ast,\tau) + g^{(1)}(z_x+z_\ast,\tau)\right)\right]\,.
\end{align}
It is easy to check that the class of functions defined in this way satisfies the four properties outlined above: First, there is a one-to-one map between the kernels $\Psi_{\pm n}$ and the functions $g^{(n)}(z_x\pm z_c,\tau)$. Since the latter define the basis of eMPLs $\widetilde{\Gamma}$, there is a one-to-one map between the functions $\cE_4$ and $\widetilde{\Gamma}$, and so the iterated integrals in eq.~\eqref{eq:cE4_def} define a basis for the space of all eMPLs. Second, since the coefficients in eq.~\eqref{eq:Psi_to_gh} are all $\pm1$, the functions $\cE_4$ can be written as a $\mathbb{Q}$-linear combination of $\widetilde{\Gamma}$ functions, and so the $\cE_4$ functions are pure. Third, it is easy to see that eq.~\eqref{eq:Psi_to_gh} has definite parity under changing the sign of $z_x$. Hence, the $\cE_4$ functions define a pure basis of eMPLs with definite parity. Finally, the term proportional to a Kronecker $\delta$ is conventional, and added so that (cf. eq.~\eqref{eq:psi_to_g})
\beq
dx\,\Psi_1(c,x,\vec a) = \frac{dx}{x-c}\,,\qquad c\neq \infty\,.
\eeq
In this way we make manifest that ordinary MPLs are a subset of eMPLs,
\beq\label{eq:cE4_to_G}
\cEf{1 & \ldots & 1}{c_1 & \ldots& c_k}{x}{\vec{a}} = G(c_1,\ldots,c_k;x)\,,\qquad c_i\neq \infty\,.
\eeq
Given the properties that the iterated integrals in eq.~\eqref{eq:cE4_def} fulfil, we argue that this class of functions is very well suited to express Feynman integrals that can be written in terms of eMPLs. We will illustrate this on several non-trivial elliptic Feynman integrals in Section~\ref{sec:applications}. In the remainder of this section we study in more detail the properties of the $\cE_4$ functions.

\subsection{Integration kernels defining pure eMPLs}

So far we have defined the kernels $\Psi_{\pm n}$ in eq.~\eqref{eq:Psi_to_gh} only implicitly through their relationship to the coefficients of the Eisenstein-Kronecker series. In ref.~\cite{Broedel:2017kkb} it was shown that there is a one-to-one map between the $g^{(n)}$ functions and the kernels $\psi_{\pm n}$ defined in Section~\ref{sec:empls}. Using the results of ref.~\cite{Broedel:2017kkb} we can give an explicit representation of the kernels that appear in eq.~\eqref{eq:cE4_def}. We present here explicitly the formulas up to $n=1$, and the corresponding formulas for $n=2$ are given in Appendix~\ref{app:kernels_2}. The extension to higher values of $n$ is straightforward. For $n=0$, we find
\beq\label{eq:pure_psi0}
\Psi_0(0,x,\vec a) = \frac{1}{\omega_1}\,\psi_0(0,x,\vec a)= \frac{c_4}{\omega_1\,y}\,.
\eeq
For $n=1$, we have (with $c\neq \infty$)
\begin{align}
\nonumber \Psi_1(c,x,\vec a) &\,= \psi_1(c,x,\vec a)= \frac{1}{x-c}\,, \\
\label{eq:pure_psi1}\Psi_{-1}(c,x,\vec a) &\,= \psi_{-1}(c,x,\vec a) + Z_4(c,\vec a)\,\psi_0(0,x,\vec a) = \frac{y_c}{y(x-c)} + Z_4(c,\vec a)\,\frac{c_4}{y}\,,\\
\nonumber\Psi_{1}(\infty,x,\vec a) &\,= -\psi_{1}(\infty,x,\vec a) = -Z_4(x,\vec a)\,\frac{c_4}{y}\,,\\
\nonumber\Psi_{-1}(\infty,x,\vec a) &\,= \psi_{-1}(\infty,x,\vec a) - \left[\frac{a_1}{c_4}+ 2 G_{\ast}(\vec a)\right]\,\psi_0(0,x,\vec a) = \frac{x}{y}   -\frac{1}{y} \left[{a_1}+ 2c_4\, G_{\ast}(\vec a)\right]\,.
\end{align}
The quantity $G_{\ast}(\vec a)$ in the last equation corresponds to the image of $z_\ast$ under the function $g^{(1)}$,
\beq\label{eq:G_infty_def}
G_{\ast}(\vec a) \equiv \frac{1}{\omega_1}\,g^{(1)}(z_{\ast},\tau)\,.
\eeq
In eq.~\eqref{eq:zstar} we have seen that $z_{\ast}$ can be expressed in terms of elliptic integrals of the first kind. Similarly, it is possible to derive a closed analytic expression for $G_{\ast}(\vec a)$ in terms of elliptic integrals of the first and second kind. In the following we discuss only the case where the branch points are real and ordered in the natural way. In this case we can use eq.~\eqref{eq:zstar} to write $z_{\ast}$ in terms of the function $\cZ_{\ast}(\alpha,\lambda)$.  Performing exactly the same steps as in the derivation of eq.~\eqref{eq:zstar} (see Appendix~\ref{app:zstar}), we find
\beq\bsp\label{eq:G_infty_KE}
G_{\ast}(\vec a)&\,=
\frac{1}{\omega_1}\,g^{(1)}\left(\cZ_{\ast}(\alpha,\lambda),\tau\right)=\lim_{x\to\infty}\left[\frac{y}{2c_4\,(x-a_1)} - \frac{1}{2}\,Z_4(x,\vec a)\right]\\
&\, = \left(\frac{2  \eta_1}{ \omega_1}-\frac{\lambda }{3}+\frac{2}{3}\right) \textrm{F}\!\left(\sqrt{\alpha}|\lambda \right)-\textrm{E}\!\left(\sqrt{\alpha }|\lambda \right)+\sqrt{\frac{\alpha  (\alpha  \lambda -1)}{\alpha -1}}\,.
\esp\eeq
We stress that this relation only holds in the case where the branch points are ordered in the natural way and the branches of the square root are chosen according to eq.~\eqref{eq:rsigns}. Just like in the case of the relation between $z_{\ast}$ and $\cZ_{\ast}(\alpha,\lambda)$, the relation remains true in other cases up to a sign and up to complex conjugation, cf. Appendix~\ref{app:zstar}.

We see that the price to pay to have integration kernels that define pure functions is that the kernels involve the functions $Z_4$ and $G_{\ast}$. While in general these functions are transcendental, in applications these functions can often be expressed in terms of algebraic quantities, thereby simplifying greatly the analytic structure of the integration kernels $\Psi_n$. Let us illustrate this on the example of the function $G_{\ast}$. We focus again on the region where the branch points are real and ordered in the natural way and refer Appendix~\ref{app:zstar} for the other cases. We start from eq.~\eqref{eq:zstar}, which relates $z_{\ast}$ and $\cZ_{\ast}(\alpha,\lambda)$. In applications, one often encounters the situation that $z_{\ast}$ takes the particularly simple form
\beq\label{eq:zstar_rational}
z_{\ast}  = a+b\,\tau(\lambda) = a + b \tau(\lambda) = a+ib\,\frac{\textrm{K}(1-\lambda)}{\textrm{K}(\lambda)}\,,\qquad a,b\textrm{ constant}.
\eeq
Often we even have $a,b\in\mathbb{Q}$, in which case $z_{\ast}$ is a rational point (torsion point) on the elliptic curve. Equating eq.~\eqref{eq:zstar} and~\eqref{eq:zstar_rational}, we find,
\beq\label{eq:cZ_tau}
\cZ_{\ast}(\alpha,\lambda)  = \frac{1}{2} - \frac{\textrm{F}(\sqrt{\alpha} | \lambda)}{ 2\, \textrm{K}(\lambda)} = a+b\,\tau(\lambda)\,.
\eeq
We see that the left-hand side depends on both $\alpha$ and $\lambda$, while the right-hand side depends only on $\lambda$.
This implies that in eq.~\eqref{eq:zstar} $\alpha$ and $\lambda$ cannot be independent,
but $\alpha=\alpha(\lambda)$ must be a function of $\lambda$. 
In physics applications this situation is encountered frequently, because the branch points, and thus both $\alpha$ and $\lambda$, are usually (algebraic) functions of the external kinematic data (Mandelstam invariants and masses), so that $\alpha$ and $\lambda$ are not independent and, at least locally, we can express $\alpha$ in terms of $\lambda$. Differentiating eq.~\eqref{eq:cZ_tau} with respect to $\lambda$ and using eq.~\eqref{eq:zstar}, we find
 \begin{align}
 b\, \frac{d \tau}{d \lambda} &= -\, \frac{d}{d\lambda}\frac{\textrm{F}(\sqrt{\alpha(\lambda)} | \lambda)}{ 2\, \textrm{K}(\lambda)}\,. \label{eq:2}
 \end{align}
As it is well known, the derivative of ${\rm F}(\sqrt{\alpha(\lambda)}| \lambda)$ involves the function ${\rm E}(\sqrt{\alpha(\lambda)}| \lambda)$,
such that by working out the derivative we can invert eqs.~(\ref{eq:cZ_tau}) and (\ref{eq:2}) and express ${\rm F}(\sqrt{\alpha(\lambda)}| \lambda)$
and ${\rm E}(\sqrt{\alpha(\lambda)}| \lambda)$ in terms of $\tau$ and its derivative. Substituting these results into
eq.~\eqref{eq:G_infty_def}, we are left with
\begin{align}
G_{\ast}(\vec a) &= 
-\frac{\lambda  (\lambda -1) \alpha '(\lambda )}{\sqrt{ \alpha (1-\alpha )  (1-\alpha  \lambda )}}-
\frac{\alpha  (\lambda -1)}{\sqrt{\alpha (1-\alpha ) (1-\alpha \lambda )}}
-2 \,b\,  \lambda  (\lambda -1) \omega_1 \tau '(\lambda )
\end{align}
where $'$ indicates the derivative with respect to $\lambda$, and we suppressed the dependence of $\alpha$ on $\lambda$.
It is very easy to compute $\tau'(\lambda)$ as
\begin{align}
\tau'(\lambda) &= i\,\frac{d}{d\lambda} \frac{\textrm{K}(1-\lambda)}{\textrm{K}(\lambda)} = \frac{i \pi }{(\lambda -1) \lambda  \, \omega_1^2}
\end{align}
such that the expression above becomes
\begin{align}
G_{\ast}(\vec a) &= \frac{(1-\lambda)\left[\lambda\,  \alpha '(\lambda ) + \alpha\right]}{\sqrt{ \alpha (1-\alpha )  (1-\alpha  \lambda )}}
-b\,  \frac{ 2\pi i }{ \omega_1} \,.\label{eq:G_start_algebraic}
\end{align}
Let us make some comments about eq.~\eqref{eq:G_start_algebraic}. First, we stress that eq.~\eqref{eq:G_start_algebraic} is only valid when the branch points are real and the branches of the square root are chosen according to eq.~\eqref{eq:rsigns}. In other cases the formula holds up to a sign and complex conjugation, see Appendix~\ref{app:zstar}. Second, eq.~\eqref{eq:G_start_algebraic} assumes that $\alpha$ and $\lambda$ are not independent, and that in addition $z_{\ast}$ takes the special form in eq.~\eqref{eq:zstar_rational}. 
Once the exact relation between $\alpha$ and $\lambda$ is known (which of course depends on the problem considered), eq.~\eqref{eq:G_start_algebraic} becomes explicit and can be used to derive the expression for $G_{\ast}(\vec a)$. In physics applications, both $\alpha$ and $\lambda$ are usually algebraic functions of the external kinematics, in which case $G_{\ast}(\vec a)$ reduces to an (explicitly computable) algebraic function of the external kinematic data, up to the term proportional to $i\pi/\omega_1$. We will see 
an explicit example of this in the next section, when we discuss results for some Feynman integrals that evaluate
to pure combinations of elliptic polylogarithms. 


\subsection{Properties of pure eMPLs}
Before we discuss examples of Feynman integrals that can be expressed in terms of the pure basis of eMPLs defined in the previous subsection, we summarise here some of their properties. Most of these properties are inherited from the corresponding properties of the $\textrm{E}_4$ and $\widetilde{\Gamma}$ functions, but we collect them here for completeness.

\paragraph{Shuffle algebra.} Just like ordinary MPLs (and iterated integrals in general), the $\cE_4$ functions form a shuffle algebra,
\beq\label{eq:shuffle_cE4}
\cE_4(A_1 \cdots A_k;x,\vec a)\,\cE_4(A_{k+1}\cdots A_{k+l};x,\vec a) = \sum_{\sigma\in \Sigma(k,l)}\cE_4(A_{\sigma(1)} \cdots A_{\sigma(k+l)};x,\vec a)\,,
\eeq
with $A_i=\left(\begin{smallmatrix}n_i\\ c_i\end{smallmatrix}\right)$.

\paragraph{Rescaling of arguments.} Just like ordinary MPLs, the $\cE_4$ functions are invariant under a simultaneous rescaling of the arguments (cf. eq.~\eqref{eq:rescaling}),
\beq
\cEf{n_1 & \ldots & n_k}{p\,c_1 & \ldots& p\,c_k}{p\,x}{p\,\vec{a}} = \cEf{n_1 & \ldots & n_k}{c_1 & \ldots& c_k}{x}{\vec{a}}\,,\qquad p,c_k\neq 0\,.
\eeq

\paragraph{Unipotency, symbols and coaction.} The $\cE_4$ functions are pure linear combinations of $\widetilde{\Gamma}$ functions. The latter are unipotent~\cite{BrownLevin,Broedel:2018iwv}, i.e., they satisfy a differential equation without homogeneous term (cf. eq.~\eqref{eq:gamma_differential}). It immediately follows that the $\cE_4$ functions are also unipotent. We can then apply the construction of ref.~\cite{Brown:coaction} and define a notion of symbols and a coaction on $\cE_4$ functions, similar to the coaction on the $\widetilde{\Gamma}$ functions introduced in ref.~\cite{Broedel:2018iwv}.
For example, we have
\begin{align}
\nonumber\Delta&\left(\cEf{0&-1}{0&c}{x}{\vec{a}}\right) = \cEf{0&-1}{0&c}{x}{\vec{a}}\otimes 1 - \cEf{-1}{\infty}{x}{\vec{a}}\otimes \left[dz_c\right] + \cEf{-1}{c}{x}{\vec{a}}\otimes \left[dz_x\right]\\
& - \cEf{1}{c}{x}{\vec{a}}\otimes \left[dz_c\right] + \cEf{-2}{\infty}{x}{\vec{a}}\otimes \left[\frac{d\tau}{2\pi i}\right] - \cEf{0}{0}{x}{\vec{a}}\otimes \left[\omega^{(1)}(z_0-z_c,\tau)\right]\\
\nonumber&+\cEf{0}{0}{x}{\vec{a}}\otimes \left[\omega^{(1)}(-z_c,\tau)\right] + \cEf{0}{0}{x}{\vec{a}}\otimes \left[\omega^{(1)}(z_c,\tau)\right] \\
\nonumber&+\cEf{0}{0}{x}{\vec{a}}\otimes \left[\omega^{(1)}(z_c+z_0,\tau)\right]
+1\otimes S\,,
\end{align}
where $S$ denotes the symbol of $\cEf{0&-1}{0&c}{x}{\vec{a}}$,
\begin{align}
\nonumber&S = \big[dz_0 - dz_x\big|\omega ^{(1)}\left(z_0-z_c,\tau \right)\big] +\big[ dz_x-dz_0\big|\omega ^{(1)}\left(-z_c,\tau \right)\big] +\big[dz_0 - dz_x\big|\omega ^{(1)}\left(z_c,\tau \right)\big]\\
\nonumber& +\big[ dz_x -dz_0\big|\omega ^{(1)}\left(z_c+z_0,\tau \right)\big] +\big[ \omega ^{(1)}\left(z_0-z_c,\tau \right)\big|dz_0-dz_x\big] +\big[ \omega ^{(1)}\left(z_c+z_0,\tau \right)\big|dz_x+dz_c\big]\\
\nonumber& +\big[ \omega ^{(1)}\left(z_x-z_c,\tau \right)\big|dz_x-dz_c\big] -\big[ \omega ^{(1)}\left(z_c+z_x,\tau \right)\big|dz_c-dz_x\big] +\big[ \omega ^{(2)}\left(z_x-z_c,\tau \right)\big|\frac{d\tau}{2\pi i}\big] \\
&-\big[ \omega ^{(2)}\left(z_c+z_x,\tau \right)\big|\frac{d\tau}{2\pi i}\big] -\big[ \omega ^{(2)}\left(z_0-z_c,\tau \right)\big|\frac{d\tau}{2\pi i}\big] +\big[ \omega ^{(2)}\left(z_c+z_0,\tau \right)\big|\frac{d\tau}{2\pi i}\big]\,.
\end{align}
Note that via eq.~\eqref{eq:cE4_to_G} we can interpret every ordinary MPL as an eMPL. Very importantly, the definition of the symbol in the elliptic case agrees with the definition of the symbol of ordinary MPLs~\cite{ChenSymbol,Goncharov:2009tja,Brown:2009qja,Goncharov:2010jf,Duhr:2011zq}.

\paragraph{Closure under integration.} Consider the algebra $\cA_4$ generated by functions of the form
\beq
R(x,y) \, Z_4(x,\vec a)^m\,\cEf{n_1 & \ldots & n_k}{c_1 & \ldots& c_k}{x}{\vec{a}}\,,
\eeq
where $m\in\mathbb{N}$ and $R(x,y)$ is a rational function, with $y^2=P_4(x)$ and we assume that $\vec a$ and the $c_i$ are independent of $x$. Then every element of $\cA_4$ has a primitive with respect to $x$. This follows immediately from the corresponding statement with $\cE_4$ replaced by $\textrm{E}_4$, which was proved in ref.~\cite{Broedel:2017kkb} (see also ref.~\cite{BrownLevin}). The computation of the primitive is algorithmic, and can be done using the techniques of ref.~\cite{Broedel:2017kkb}.

\paragraph{Relationship to iterated integrals of Eisenstein series.} In ref.~\cite{Broedel:2018iwv} it was shown that whenever the arguments of $\gamtt{n_1 &\ldots& n_k}{z_1 & \ldots & z_k}{z_{k+1}}{\tau}$ are rational points\footnote{In the mathematics literature, these points are known as \emph{torsion points} of the elliptic curve.} of the form $z_i = \frac{r_i}{N} + \tau\frac{s_i}{N}$, $r_i,s_i,N\in\mathbb{Z}$ and $N>0$, then it can be expressed in terms of iterated integrals of Eisenstein series for $\Gamma(N) = \left\{\gamma\in SL(2,\mathbb{Z}):\gamma = \left(\begin{smallmatrix}1&0\\0&1\end{smallmatrix}\right)\!\!\mod N\right\}$. A spanning set for Eisenstein series for $\Gamma(N)$ is given by~\cite{Broedel:2018iwv,ZagierBonn}
\beq
h^{(n)}_{N,r,s}(\tau) =-\!\!\!\!\!\!\!\sum_{\substack{(\alpha,\beta)\in \mathbb{Z}^2\\ (\alpha,\beta)\neq(0,0)}}\frac{e^{2\pi i(s\alpha-r\beta)/N}}{(\alpha+\beta\tau)^{2n}}\,.
\eeq
The iterated integrals one needs to consider are thus
\beq\bsp\label{eq:I_Eisenstein_def}
I\left(\begin{smallmatrix} n_1& N_1\\ r_1& s_1\end{smallmatrix}\big|\ldots\big|\begin{smallmatrix} n_k& N_k\\ r_k& s_k\end{smallmatrix};\tau\right) = \int_{i\infty}^{\tau}d\tau'\,h_{N_1,r_1,s_1}^{(n_1)}(\tau')\,I\left(\begin{smallmatrix} n_2& N_2\\ r_2& s_2\end{smallmatrix}\big|\ldots\big|\begin{smallmatrix} n_k& N_k\\ r_k& s_k\end{smallmatrix};\tau'\right)\,.
\esp\eeq
By convention, we set $h_{0,0,0}^{(0)}(\tau)\equiv 1$. In general, these integrals require regularisation~\cite{Brown:mmv}. The weight of $I\left(\begin{smallmatrix} n_1& N_1\\ r_1& s_1\end{smallmatrix}\big|\ldots\big|\begin{smallmatrix} n_k& N_k\\ r_k& s_k\end{smallmatrix};\tau\right) $ is defined as $\sum_in_i$. Since every $\cE_4$ function can be written as a pure linear combination of $\widetilde{\Gamma}$'s, we conclude that a similar statement holds for $\cE_4$ functions. More precisely, consider the function $\cEf{n_1 & \ldots & n_k}{c_1 & \ldots& c_k}{x}{\vec{a}}$. If all of its arguments, including the point at infinity and the base point, are mapped to rational points on the torus by Abel's map (i.e., $z_0$, $z_{\ast}$, $z_x$ and all of the $z_{c_i}$ are rational points), then we can express $\cEf{n_1 & \ldots & n_k}{c_1 & \ldots& c_k}{x}{\vec{a}}$ as a pure linear combination (of uniform weight $\sum_i|n_i|$) of the iterated integrals defined in eq.~\eqref{eq:I_Eisenstein_def}.

\paragraph{Value at the infinite cusp.} For $\tau\to i\infty$, $\cEf{n_1 & \ldots & n_k}{c_1 & \ldots& c_k}{x}{\vec{a}}$ always reduces to a pure  combination of ordinary MPLs of weight $\sum_in_i$ (provided that the limit exists). This follows immediately from analysing how the coefficients of the Eisenstein-Kronecker series behave as $\tau\to i\infty$. In particular, they admit the Fourier expansions~\cite{Broedel:2014vla,Broedel:2015hia,Broedel:2017jdo}
\beq\bsp\label{eq:g_q_exp}
g^{(1)}(z,\tau) &= \pi  \cot(\pi z) + 4\pi \sum_{m=1}^{\infty}  \sin(2\pi m z) \sum_{n=1}^{\infty} q^{mn}\,,\\
g^{(k)}(z,\tau) \Big|_{k=2,4,\ldots} &=  - 2\zeta_k - 2 \frac{ (2\pi i)^k }{(k-1)!} \sum_{m=1}^{\infty}  \cos(2\pi m z) \sum_{n=1}^{\infty}n^{k-1} q^{mn} \,, \\
g^{(k)}(z,\tau) \Big|_{k=3,5,\ldots}&= - 2i \frac{ (2\pi i)^k }{(k-1)!} \sum_{m=1}^{\infty}  \sin(2\pi m z) \sum_{n=1}^{\infty}n^{k-1} q^{mn}\,,
\esp\eeq
with $q = \exp(2\pi i\tau)$.

\paragraph{Regularisation.} In the case where $(n_k,c_k) = (\pm1,0)$ the integral in eq.~\eqref{eq:cE4_def} is divergent and requires regularisation. This is similar to the case of ordinary MPLs, where the naive iterated integral representation for $G(0,\ldots,0;x)$ diverges and instead one gives a special definition, cf. eq.~\eqref{eq:GLog}. In the case of eMPLs, we have a divergence whenever $(n_k,c_k) = (\pm1,0)$, and so we need a special definition for the cases $\cEf{n_1 & \ldots & n_k}{0&\ldots&0}{x}{\vec a}$, with $n_i=\pm1$. We define, for $A_i=\left(\begin{smallmatrix}\pm1\\0\end{smallmatrix}\right)$,
\begin{align}\label{eq:eMPL_reg}
\cE_4(A_1\ldots A_k;x;\vec a) = \frac{1}{k!}\log^kx + &\sum_{l=0}^k\sum_{m=1}^l\sum_{\sigma}\frac{(-1)^{l+m}}{(k-l)!}\log^{k-l}x\\
\nonumber&\times\cE_4^{\textrm{R}}\left(A^{(m)}_{\sigma(1)}\ldots A^{(m)}_{\sigma(m-1)} A^{(m)}_{\sigma(m+1)}\ldots A^{(m)}_{\sigma(l)}\Big|A_m;x;\vec a\right)\,,
\end{align}
where the third sum runs over all shuffles $\sigma\in\Sigma(m-1,l-m)$ and $A_i^{(m)} = A_i$ if $i<m$ and $A_i^{(m)} = \left(\begin{smallmatrix}1\\0\end{smallmatrix}\right)$ otherwise. The $\cE_4^{\textrm{R}}$ are iterated integrals with suitable subtractions to render the integrations finite,
\beq
\!\!\!\cE_4^{\textrm{R}}\left(\begin{smallmatrix} n_1&\ldots & n_k \\ 0 &\ldots &0\end{smallmatrix}|\begin{smallmatrix} n_a \\ 0 \end{smallmatrix};x;\vec a\right) 
= \int_0^xdt_1\Psi_{n_1}(0,t_1)\int_{0}^{t_1}\ldots \int_{0}^{t_{k-1}}dt_k\left(\Psi_{n_a}(0,t_k) - \Psi_{1}(0,t_k)\right)\,.
\eeq
For example, we have
\begin{align}\label{eq:regularisation_example}
\nonumber\cEf{-1}{0}{x}{\vec a} &\,= \log x + \cE_4^{\textrm{R}}\left(|\begin{smallmatrix} -1 \\ 0 \end{smallmatrix};x;\vec a\right) \,,\\
\cEf{1&-1}{0&0}{x}{\vec a} &\,= \frac{1}{2}\log^2 x + \cE_4^{\textrm{R}}\left(\begin{smallmatrix} 1\\ 0 \end{smallmatrix}|\begin{smallmatrix}-1\\ 0 \end{smallmatrix};x;\vec a\right)\,,\\
\nonumber\cEf{-1&1}{0&0}{x}{\vec a} &\,= \frac{1}{2}\log^2 x + \log x\,\cE_4^{\textrm{R}}\left(|\begin{smallmatrix} -1 \\ 0 \end{smallmatrix};x;\vec a\right) -  \cE_4^{\textrm{R}}\left(\begin{smallmatrix} 1\\ 0 \end{smallmatrix}|\begin{smallmatrix}-1\\ 0 \end{smallmatrix};x;\vec a\right)\,,\\
\nonumber\cEf{-1&-1}{0&0}{x}{\vec a} &\,= \frac{1}{2}\log^2 x + \log x\,\cE_4^{\textrm{R}}\left(|\begin{smallmatrix} -1 \\ 0 \end{smallmatrix};x;\vec a\right) -  \cE_4^{\textrm{R}}\left(\begin{smallmatrix} -1\\ 0 \end{smallmatrix}|\begin{smallmatrix}-1\\ 0 \end{smallmatrix};x;\vec a\right)+\cE_4^{\textrm{R}}\left(\begin{smallmatrix} 1\\ 0 \end{smallmatrix}|\begin{smallmatrix}-1\\ 0 \end{smallmatrix};x;\vec a\right)\,,
\end{align}
with
\beq\bsp
\cE_4^{\textrm{R}}\left(|\begin{smallmatrix} -1\\ 0 \end{smallmatrix};x;\vec a\right) &\,= \int_0^xdt\left(\Psi_{-1}(0,t) - \Psi_1(0,t)\right)\,,\\
\cE_4^{\textrm{R}}\left(\begin{smallmatrix} \pm1\\ 0 \end{smallmatrix}|\begin{smallmatrix}-1\\ 0 \end{smallmatrix};x;\vec a\right) &\,= \int_0^xdt_1\,\Psi_{\pm1}(0,t_1)\int_0^{t_1}dt_2\,\left(\Psi_{-1}(0,t_2) - \Psi_1(0,t_2)\right)\,.
\esp\eeq
While manifestly finite, the form of the regulated eMPLs proposed in eq.~\eqref{eq:eMPL_reg} seems rather ad hoc. This is not so, and the form is in fact dictated by requiring the  following natural properties:
\begin{enumerate}
\item The regularisation of eMPLs is consistent with the regularisation of ordinary MPLs, i.e., eq.~\eqref{eq:cE4_to_G} still holds after regularisation.
\item The regularisation preserves the shuffle algebra structure, i.e., eq.~\eqref{eq:shuffle_cE4} still holds after regularisation.
\item The regularisation preserves the derivative with respect to $x$, i.e., eq.~\eqref{eq:cE4_derivative} still holds after regularisation.
\item Since $\Psi_{-1}(0,x) = \frac{1}{x} + \ord(x^0)$, the regulated value for $\cEf{n_1&\ldots&n_k}{0&\ldots&0}{x}{\vec a}$ with $n_i=\pm1$ has a logarithmic singularity for $x=0$,
\beq
\cEf{n_1&\ldots&n_k}{0&\ldots&0}{x}{\vec a} \sim \cEf{1&\ldots&1}{0&\ldots&0}{x}{\vec a} = \frac{1}{k!}\log^kx\,,\quad\textrm{ if } x\to 0\,.
\eeq
\end{enumerate}
Equation~\eqref{eq:eMPL_reg} is in fact a special case of a more general construction, cf.~ref.~\cite{Brown:mmv}. When combined with the aforementioned requirements, this construction essentially fixes the form of the regulated eMPLs to eq.~\eqref{eq:eMPL_reg}. We emphasise that it is non-trivial to find a form for the regularisation consistent with the above requirements. For example, though not obvious from the expressions in eq.~\eqref{eq:regularisation_example}, one can check that they satisfy the shuffle identities (cf. eq.~\eqref{eq:shuffle_cE4}),
\beq\bsp
\cEf{-1&-1}{0&0}{x}{\vec a} &\,= \frac{1}{2}\cEf{-1}{0}{x}{\vec a}^2\,,\\
\cEf{-1}{0}{x}{\vec a} \cEf{1}{0}{x}{\vec a} &\, = \cEf{-1&1}{0&0}{x}{\vec a} + \cEf{1&-1}{0&0}{x}{\vec a}\,,
\esp\eeq
and the differential equation in eq.~\eqref{eq:cE4_derivative},
\beq\bsp
\partial_x\cEf{-1&-1}{0&0}{x}{\vec a} &\,= \Psi_{-1}(0,x)\,\cEf{-1}{0}{x}{\vec a}\,,\\
\partial_x\cEf{-1&1}{0&0}{x}{\vec a} &\,= \Psi_{-1}(0,x)\,\cEf{1}{0}{x}{\vec a}\,,\\
\partial_x\cEf{1&-1}{0&0}{x}{\vec a} &\,= \Psi_{1}(0,x)\,\cEf{-1}{0}{x}{\vec a}\,.
\esp\eeq


\section{Some two-loop examples}
\label{sec:applications}

\subsection{Elliptic Feynman integrals and elliptic purity}

\begin{figure}[!t]
\begin{center}
\includegraphics[scale=0.8]{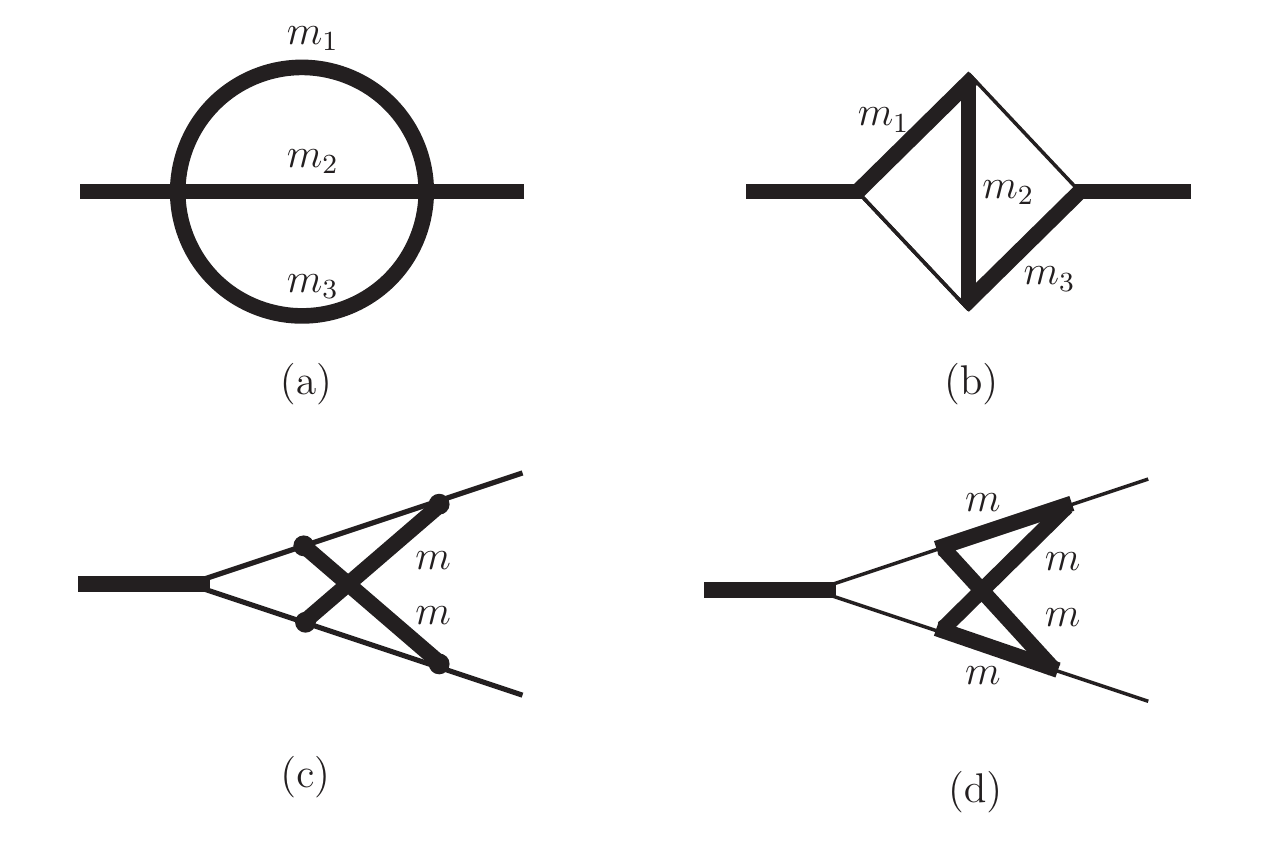}
\caption{\label{fig:examples}The collection of two-loop Feynman integrals of uniform weight that we have evaluated analytically. Thick lines denote massive propagators.}
\end{center}
\end{figure}

In this section we present some explicit examples of two-loop Feynman integrals that evaluate to pure eMPLs. More precisely, we have computed the two-loop integrals in fig.~\ref{fig:examples}. The details of the computation of the integrals will be presented elsewhere~\cite{trianglepaper}. All integrals can be evaluated in terms of eMPLs, and in all cases we find that these integrals evaluate to pure functions of uniform weight, up to an overall semi-simple factor. Before we discuss some special cases in detail in subsequent sections and illustrate some of the features, we give a brief summary of the integrals shown in fig.~\ref{fig:examples}.
\begin{enumerate}
\item The two-loop sunrise integral (fig.~\ref{fig:examples}a) in $D=2-2\eps$ dimensions was evaluated analytically in terms of elliptic generalisations of polylogarithms in ref.~\cite{Bloch:2013tra,Adams:2013kgc,Adams:2013nia,Adams:2014vja,Adams:2015gva,Adams:2015ydq,Remiddi:2016gno,Remiddi:2017har,Broedel:2017siw}. We find that it evaluates to a pure function of weight two, in agreement with the corresponding result for massless propagators, which can be expressed in terms of ordinary MPLs. We will discuss the equal-mass case in detail in Section~\ref{sec:pure_sunrise}.
\item The two-loop kite integral (fig.~\ref{fig:examples}b) in $D=4-2\eps$ dimensions was first considered in ref.~\cite{Sabry}. More recently it was  evaluated in the equal-mass case in terms of elliptic generalisations of MPLs in ref.~\cite{Remiddi:2016gno,Adams:2016xah}. We have computed the case of a kite integral with three different masses, and we find that it can expressed in terms of a pure combination of ordinary MPLs and eMPLs of uniform weight three~\cite{trianglepaper}. The equal-mass case will be discussed in detail in Section~\ref{sec:pure_kite}.
\item The non-planar three-point functions (fig.~\ref{fig:examples}c\&d) in $D=4-2\eps$ were first considered in refs.~\cite{Aglietti:2007as,vonManteuffel:2017hms}, respectively. We have evaluated them in terms of a pure combination of eMPLs of uniform weight four~\cite{trianglepaper}, in agreement with the case of massless propagators, which is known to give rise to non-elliptic functions of uniform weight four~\cite{Gonsalves:1983nq,vanNeerven:1985xr,Kramer:1986sr,Gehrmann:2005pd}. We will illustrate this in detail in Section~\ref{sec:pure_triangle} on the example of the graph shown in fig.~\ref{fig:examples}d.
\end{enumerate}
We believe that these examples give strong evidence that there is a natural way to extend the notion of Feynman integrals of uniform weight beyond the case of ordinary MPLs, and this notion of weight agrees with the weight known from the corresponding non-elliptic cases. More applications, including to four-point functions, will follow in separate publications. In the remainder of this section we analyse the three cases above in more detail.

\subsection{The two-loop sunrise integral}
\label{sec:pure_sunrise}
We start by rewriting the two-loop equal-mass sunrise integral in eq.~\eqref{eqn:S111_result} in terms of pure eMPLs. In order to do this, we invert the relations in eq.~\eqref{eq:pure_psi0} and eq.~\eqref{eq:pure_psi1} to express $\psi_{0}$ and $\psi_{\pm1}$ in terms of $\Psi_{0}$ and $\Psi_{\pm1}$. We find
\beq\bsp\label{eq:pure_sunrise}
S_{1}&(p^2,m^2) = -\frac{\omega_1}{(p^2+m^2)\,c_4}\,T_1(p^2,m^2)\,,
\esp\eeq
with
\beq\label{eq:T1_def}
T_1(p^2,m^2) = \left(\frac{m^2}{-p^2}\right)^{-2\eps}\,\left[T_1^{(0)}+\eps\,T_1^{(1)}+\ord(\eps^2)\right]\,,
\eeq
and
\begin{align}\label{eq:sunrise_1_T}
T_1^{(0)}&\,=2\cEf{0&-1}{0&\infty}{1}{\vec a} + \cEf{0&-1}{0&0}{1}{\vec a} + \cEf{0&-1}{0&1}{1}{\vec a}\,,\\
\nonumber T_1^{(1)}&\,=-4 \cEf{0 & 1 & -1}{0 & a_3 & \infty }{1}{\vec {a}}-4 \cEf{0 & 1 & -1}{0 & a_1 & \infty }{1}{\vec {a}}-4 \cEf{0 & 1 & -1}{0 & a_4 & \infty }{1}{\vec {a}}-4 \cEf{0 & 1 & -1}{0 & a_2 & \infty }{1}{\vec {a}}\\
\nonumber&\,-2 \cEf{0 & 1 & -1}{0 & a_3 & 0}{1}{\vec {a}}-2 \cEf{0 & 1 & -1}{0 & a_3 & 1}{1}{\vec {a}}-2 \cEf{0 & 1 & -1}{0 & a_1 & 0}{1}{\vec {a}}-2 \cEf{0 & 1 & -1}{0 & a_1 & 1}{1}{\vec {a}}\\
\nonumber&\,-2 \cEf{0 & 1 & -1}{0 & a_4 & 0}{1}{\vec {a}}-2 \cEf{0 & 1 & -1}{0 & a_4 & 1}{1}{\vec {a}}-2 \cEf{0 & 1 & -1}{0 & a_2 & 0}{1}{\vec {a}}-2 \cEf{0 & 1 & -1}{0 & a_2 & 1}{1}{\vec {a}}\\
\nonumber&\,+2 \cEf{0 & -1 & 1}{0 & \infty  & 0}{1}{\vec {a}}+2 \cEf{0 & -1 & 1}{0 & \infty  & 1}{1}{\vec {a}}+6 \cEf{0 & 1 & -1}{0 & 0 & \infty }{1}{\vec {a}}+6 \cEf{0 & 1 & -1}{0 & 1 & \infty }{1}{\vec {a}}\\
\nonumber&\,-2 \cEf{0 & -1 & 1}{0 & 0 & 0}{1}{\vec {a}}-2 \cEf{0 & -1 & 1}{0 & 0 & 1}{1}{\vec {a}}-2 \cEf{0 & -1 & 1}{0 & 1 & 0}{1}{\vec {a}}-2 \cEf{0 & -1 & 1}{0 & 1 & 1}{1}{\vec {a}}\\
\nonumber&\,+6 i \pi  \cEf{0 & 0 & 1}{0 & 0 & 0}{1}{\vec {a}}+6 i \pi  \cEf{0 & 0 & 1}{0 & 0 & 1}{1}{\vec {a}}+3 \cEf{0 & 1 & -1}{0 & 0 & 0}{1}{\vec {a}}+3 \cEf{0 & 1 & -1}{0 & 0 & 1}{1}{\vec {a}}\\
\nonumber&\,+3 \cEf{0 & 1 & -1}{0 & 1 & 0}{1}{\vec {a}}+3 \cEf{0 & 1 & -1}{0 & 1 & 1}{1}{\vec {a}}+\zeta_2\, \cEf{0}{0}{1}{\vec {a}}
\,.
\end{align}
The entries in the vector $\vec a$ are given in eq.~\eqref{eqn:roots}. We work in a region where the branch points are pairwise complex conjugate, so that $P_4(x)=(x-a_1)\cdots(x-a_4)$ is positive definite for real $x$. We choose the branches of the square root in that region such that $\sqrt{P_4(x)}>0$ for all real values of $x$.

Let us discuss eq.~\eqref{eq:pure_sunrise}. First, we see that eMPLs in eq.~\eqref{eq:sunrise_1_T} have uniform weight. It is natural to assign weight one also to $\omega_1 = 2\textrm{K}(\lambda)$, because $\lim_{\lambda\to 0}\textrm{K}(\lambda) = \frac{\pi}{2}$. If we assign weight $-1$ to the dimensional regularisation parameter $\eps$, we see that all the terms in eq.~\eqref{eq:pure_sunrise} have uniform weight two. This is in agreement with the weight of the sunrise integral with at least one massless propagator, which can be expressed in terms of ordinary MPLs. Second, the prefactor multiplying the pure eMPLs in eq.~\eqref{eq:pure_sunrise} corresponds to the maximal cut of the sunrise integral computed in two dimensions,
\beq
\textrm{Cut}[S_{1}(p^2,m^2)_{|D=2}] = -\frac{\omega_1}{(p^2+m^2)\,c_4}\,.
\eeq
In other words, we find that the sunrise integral can be cast in a form which is very reminiscent of the non-elliptic case, cf. eq.~\eqref{eq:cut_bubble_MPL},
\beq
S_{1}(p^2,m^2) = \textrm{Cut}[S_{1}(p^2,m^2)_{|D=2}]\times T_1(p^2,m^2)\,,
\eeq
where $T_1$ is a pure function of uniform weight.

In ref.~\cite{Broedel:2017siw} also the master integral $S_2$ defined in eq.~\eqref{eq:sunrise_masters} was computed in terms of the eMPLs $\textrm{E}_4$. Performing the same steps as for $S_1$, we find the following representation for the three-propagator master integrals for the sunrise family,
\beq
\left(\begin{array}{c}
S_{1}(p^2,m^2) \\ S_{2}(p^2,m^2)
\end{array}\right)
=
\left(\begin{array}{cc} 
\Omega_1 & 0 \\
H_1(\eps) & -\frac{2}{m^2\,(p^2+m^2)(p^2+9m^2)\,\Omega_1}
\end{array}\right)
\left(\begin{array}{c}
T_{1}(p^2,m^2) \\ T_{2}(p^2,m^2)
\end{array}\right)\,,
\eeq
where the entries in the matrix in the right-hand side are semi-simple objects,
\beq\bsp
\Omega_1&\,=-\frac{\omega _1}{c_4 \left(m^2+p^2\right)} \textrm{~~and~~} H_1(\eps) = H_1^{(0)}+\eps\,H_1^{(1)} + \ord(\eps^2)\,,
\esp\eeq
with
\beq\bsp
H_1^{(0)}&\,=-\frac{4 c_4 \eta _1}{m^2\left(9 m^2+p^2\right)}-
\frac{\omega _1 \left(15 m^4+12 m^2 p^2+p^4\right)}{6 m^2 c_4 \left(m^2+p^2\right)^2 \left(9 m^2+p^2\right)}\,,\\
H_1^{(1)}&\,=-\frac{\omega_1 \left(45m^4+30m^2p^2+p^4\right)}{6 m^2 c_4(m^2+p^2)^2(9m^2+p^2)}\,.
\esp\eeq
We note that the function $H_1^{(0)}$ is precisely the maximal cut of the second master integral,
\beq
\textrm{Cut}[S_{2}(p^2,m^2)_{|D=2}] = -\frac{1}{3}\frac{\partial}{\partial m^2}\textrm{Cut}[S_{1}(p^2,m^2)_{|D=2}] = H_1^{(0)}\,.
\eeq
The matrix has been determined empirically through order $\eps^1\,$.
The structure of this matrix is very reminiscient of the matrix of semi-simple periods in eq.~\eqref{eq:US_matrix}. The function $T_1$ is the pure part of $S_1$ defined in eq.~\eqref{eq:T1_def}. The function $T_2$ is a new pure building block given by
\beq
\label{eq:T2_def}
T_2(p^2,m^2) = \left(\frac{m^2}{-p^2}\right)^{-2\eps}\,\left[T_2^{(0)}+\eps\,T_2^{(1)}+\ord(\eps^2)\right]\,,
\eeq
\beq\bsp\label{eq:sunrise_2_T}
T_2^{(0)} &\,= 2\cEf{-2}{\infty}{1}{\vec a} + \cEf{-2}{0}{1}{\vec a} + \cEf{-2}{1}{1}{\vec a} \,,\\
T_2^{(1)} &\,=
-2 \cEf{-2&1}{0&0}{1}{\vec{a}} + 3 \cEf{2&-1}{0&0}{1}{\vec{a}} - 2 \cEf{-2&1}{0&1}{1}{\vec{a}} + 3 \cEf{2&-1}{0&1}{1}{\vec{a}}\\
&+ 6 \cEf{2&-1}{0&\infty}{1}{\vec{a}} - 2 \cEf{-2&1}{1&0}{1}{\vec{a}} + 3 \cEf{2&-1}{1&0}{1}{\vec{a}} - 2 \cEf{-2&1}{1&1}{1}{\vec{a}}\\
&+ 3 \cEf{2&-1}{1&1}{1}{\vec{a}} + 6 \cEf{2&-1}{1&\infty}{1}{\vec{a}} - 2 \cEf{2&-1}{a_1&0}{1}{\vec{a}} - 2 \cEf{2&-1}{a_1&1}{1}{\vec{a}}\\
&- 4 \cEf{2&-1}{a_1&\infty}{1}{\vec{a}}- 2 \cEf{2&-1}{a_2&0}{1}{\vec{a}} - 2 \cEf{2&-1}{a_2&1}{1}{\vec{a}} - 4 \cEf{2&-1}{a_2&\infty}{1}{\vec{a}}\\
&- 2 \cEf{2&-1}{a_3&0}{1}{\vec{a}}- 2 \cEf{2&-1}{a_3&1}{1}{\vec{a}} - 4 \cEf{2&-1}{a_3&\infty}{1}{\vec{a}} - 2 \cEf{2&-1}{a_4&0}{1}{\vec{a}}\\
&- 2 \cEf{2&-1}{a_4&1}{1}{\vec{a}}- 4 \cEf{2&-1}{a_4&\infty}{1}{\vec{a}} + 2 \cEf{-2&1}{\infty&0}{1}{\vec{a}}
+ 2 \cEf{2&-1}{\infty&0}{1}{\vec{a}}\\
&+ 2 \cEf{-2&1}{\infty&1}{1}{\vec{a}}+ 2 \cEf{2&-1}{\infty&1}{1}{\vec{a}} + 4 \cEf{2&-1}{\infty&\infty}{1}{\vec{a}}  \\
&-\frac{i\pi}{2}\Big[
3 \cEf{-1&-1}{0&0}{1}{\vec{a}} + 2 \cEf{-1&1}{0&0}{1}{\vec{a}} - 3 \cEf{1&-1}{0&0}{1}{\vec{a}} - 2 \cEf{1&1}{0&0}{1}{\vec{a}}\\
&+ 3 \cEf{-1&-1}{0&1}{1}{\vec{a}} + 2 \cEf{-1&1}{0&1}{1}{\vec{a}} - 3 \cEf{1&-1}{0&1}{1}{\vec{a}} - 2 \cEf{1&1}{0&1}{1}{\vec{a}}\\
&+ 6 \cEf{-1&-1}{0&\infty}{1}{\vec{a}} - 6 \cEf{1&-1}{0&\infty}{1}{\vec{a}} + 3 \cEf{-1&-1}{1&0}{1}{\vec{a}} + 2 \cEf{-1&1}{1&0}{1}{\vec{a}}\\
&- 3 \cEf{1&-1}{1&0}{1}{\vec{a}} - 2 \cEf{1&1}{1&0}{1}{\vec{a}} + 3 \cEf{-1&-1}{1&1}{1}{\vec{a}} + 2 \cEf{-1&1}{1&1}{1}{\vec{a}}\\
&- 3 \cEf{1&-1}{1&1}{1}{\vec{a}} - 2 \cEf{1&1}{1&1}{1}{\vec{a}} + 6 \cEf{-1&-1}{1&\infty}{1}{\vec{a}} - 6 \cEf{1&-1}{1&\infty}{1}{\vec{a}}\\
&+ 2 \cEf{1&-1}{a_1&0}{1}{\vec{a}} + 2 \cEf{1&-1}{a_1&1}{1}{\vec{a}} + 4 \cEf{1&-1}{a_1&\infty}{1}{\vec{a}} - 4 \cEf{-1&-1}{a_2&0}{1}{\vec{a}}\\
&+ 2 \cEf{1&-1}{a_2&0}{1}{\vec{a}} - 4 \cEf{-1&-1}{a_2&1}{1}{\vec{a}} + 2 \cEf{1&-1}{a_2&1}{1}{\vec{a}} - 8 \cEf{-1&-1}{a_2&\infty}{1}{\vec{a}}\\
&+ 4 \cEf{1&-1}{a_2&\infty}{1}{\vec{a}} - 4 \cEf{-1&-1}{a_3&0}{1}{\vec{a}} + 2 \cEf{1&-1}{a_3&0}{1}{\vec{a}} - 4 \cEf{-1&-1}{a_3&1}{1}{\vec{a}}\\
&+ 2 \cEf{1&-1}{a_3&1}{1}{\vec{a}} - 8 \cEf{-1&-1}{a_3&\infty}{1}{\vec{a}} + 4 \cEf{1&-1}{a_3&\infty}{1}{\vec{a}} + 2 \cEf{1&-1}{a_4&0}{1}{\vec{a}}\\
&+ 2 \cEf{1&-1}{a_4&1}{1}{\vec{a}} + 4 \cEf{1&-1}{a_4&\infty}{1}{\vec{a}} - 2 \cEf{-1&-1}{\infty&0}{1}{\vec{a}} - 2 \cEf{-1&1}{\infty&0}{1}{\vec{a}}\\
&- 6 \cEf{1&1}{\infty&0}{1}{\vec{a}} - 2 \cEf{-1&-1}{\infty&1}{1}{\vec{a}} - 2 \cEf{-1&1}{\infty&1}{1}{\vec{a}} - 6 \cEf{1&1}{\infty&1}{1}{\vec{a}}\\
&- 4 \cEf{-1&-1}{\infty&\infty}{1}{\vec{a}}
\Big]+\pi^2\Big[
2 \cEf{0&-1}{0&0}{1}{\vec{a}} - 3 \cEf{0&1}{0&0}{1}{\vec{a}} + 2 \cEf{0&-1}{0&1}{1}{\vec{a}}\\
&- 3 \cEf{0&1}{0&1}{1}{\vec{a}} + 4 \cEf{0&-1}{0&\infty}{1}{\vec{a}}
\Big]\,.
\esp\eeq
Let us conclude this section with a few comments. First, we see that all the arguments of the elliptic polylogarithms are drawn from the set $\{0,1,\infty,a_1,\ldots,a_4\}$. For concreteness we work in the Euclidean region where the branch points in eq.~\eqref{eqn:roots} are pairwise complex conjugate to each other. Under Abel's map in eq.~\eqref{eq:Abel} the branch points are mapped to the half-periods of the elliptic curve,
\beq
z_{a_1} = 0\,,\quad z_{a_2} = \frac{\tau}{2}\,,\quad z_{a_3} = \frac{1}{2}+\frac{\tau}{2}\,,\quad z_{a_4} = \frac{1}{2}\,.
\eeq
The image of the point at infinity is given by eq.~\eqref{eq:zstar}. However, since we are considering a setup where the branch points are not real, we cannot apply eq.~\eqref{eq:zstar} out of the box, but we have to be careful about signs. Using the results from Appendix~\ref{app:zstar}, we find
\beq\label{eq:z*_sunrise}
z_{\ast} = -\cZ_{\ast}(\alpha,\lambda) = -\frac{1}{4}-\frac{\tau}{4}\,.
\eeq
Finally, a numerical evaluation of the integral in eq.~\eqref{eq:Abel} for $X=0$ or $1$ reveals that
\beq\label{eq:z01_sunrise}
z_0 = -\frac{5}{12} + \frac{\tau}{4} \quad \textrm{and}\quad z_1 = -\frac{1}{12} + \frac{\tau}{4} \,.
\eeq
We see that all the arguments of the eMPLs are mapped to rational points on the torus under Abel's map. This implies that the equal-mass sunrise integral can also be expressed in terms of iterated integrals of Eisenstein series, in agreement with ref.~\cite{Adams:2017ejb,Broedel:2018iwv}. 

Next, we note that in order to arrive at the simple pure results in eqs.~\eqref{eq:sunrise_1_T} and~\eqref{eq:sunrise_2_T}, we need to use following identities,
\begin{align}
\label{eq:sunrise_relations_1}
Z_4(0,\vec a) + Z_4(1,\vec a) &= \frac{2\pi i}{\omega_1}\,,\\
\label{eq:sunrise_relations_2}
G_{\ast}(\vec a) &= \frac{i\pi}{2\omega_1}-\frac{2\, a_1-1}{4c_4}\,.
\end{align}
These identities are necessary conditions for $S_{1}(p^2,m^2)$ to be pure up to an overall multiplicative factor. We now briefly discuss how these identities can be obtained. Let us start by analysing eq.~\eqref{eq:sunrise_relations_1}. Using eqs.~\eqref{eq:g_parity}, \eqref{eq:g_periodic} and~\eqref{eq:Z4_to_g1}, as well as~\eqref{eq:z*_sunrise} and~\eqref{eq:z01_sunrise} , we find
\beq\bsp
Z_4(0,\vec a) &\,= -\frac{1}{\omega_1}\left[g^{(1)}\left(\frac{1}{3},\tau\right)-g^{(1)}\left(\frac{1}{6} + \frac{\tau}{2},\tau\right)-2\pi i\right]\,,\\
Z_4(1,\vec a) &\,= -\frac{1}{\omega_1}\left[-g^{(1)}\left(\frac{1}{3},\tau\right)+g^{(1)}\left(\frac{1}{6} + \frac{\tau}{2},\tau\right)\right]\,,
\esp\eeq
and so eq.~\eqref{eq:sunrise_relations_1} immediately follows. Next, let us discuss eq.~\eqref{eq:sunrise_relations_2}. We see from eq.~\eqref{eq:z*_sunrise} that in the case of the equal-mass sunrise integral the point $z_{\ast}$ matches the form in eq.~\eqref{eq:zstar_rational}. Moreover, we see from eq.~\eqref{eqn:roots} that both $\alpha$ and $\lambda$ are algebraic functions of the ratio of the kinematic variables $p^2$ and $m^2$. Hence, $\alpha$ and $\lambda$ are not independent, and we can express $\alpha$ as an algebraic function of $\lambda$ (at least locally). As a consequence, we can apply eq.~\eqref{eq:G_start_algebraic}, and we immediately recover eq.~\eqref{eq:sunrise_relations_2} (we recall that we are working in a region where the branch points are complex, and so we cannot apply eq.~\eqref{eq:G_start_algebraic} out of the box,  but we need to be careful about the signs, as explained in Appendix~\ref{app:zstar}).

\subsection{The two-loop kite integral}
\label{sec:pure_kite}

As a first obvious generalisation of the sunrise graph, it is natural to consider the kite integral.
We focus here on the case of three massive and two massless propagators, see fig.~\ref{fig:examples}b. 
While the three masses could in general be different, 
the equal mass case
is relevant to compute the two-loop corrections to the electron self-energy in QED~\cite{Sabry,Remiddi:2016gno,Adams:2016xah}.

We define the kite as
\beq\bsp
K&(p^2,m_1^2,m_2^2,m_3^2)\\
 &= -\frac{e^{2\gamma_E\eps}}{\pi^D}\, \int \frac{d^D k\, d^D l}{l^2\, (k-p)^2\, (k^2-m_1^2) ((k-l)^2-m_2^2) ((l-p)^2-m_3^2)}\,,
\label{eq:kite}
\esp\eeq
and consider its Laurent expansion close to $D=4$ space-time dimensions by introducing $D=4 -2 \epsilon$.
As will be explicitly shown elsewhere, it is a straightforward exercise to express the different mass kite integral
in terms of eMPLs by direct integration over Feynman parameters. Here we will not be concerned with the details of the calculation,
but we will instead show how the kite integral can be expressed
as combination of pure functions, up to an overall rational prefactor.
As the formulas in the different mass case become quickly very lengthy due to obvious combinatorial reasons, we prefer to show
here only the result for the equal mass case, but we stress that conceptually similar formulas can be derived for the general case.
In the limit of equal masses we define

\begin{equation}
K(p^2,m^2) \equiv K(p^2,m^2,m^2,m^2) \,.
\end{equation}

For a given (real) value of the mass $m$, the kite develops a first discontinuity as $p^2 \geq m^2$ and a second one as $p^2 \geq 9 m^2$.
By introducing the dimensionless variable $z = p^2/m^2$ we see that the result is real for $z < 1$.
For the sake of simplicity, in this paper we limit ourselves to consider the region 
$0 < z <1$, i.e., $0< p^2 < m^2$.
The only elliptic curve relevant for the calculation of the kite integral is the one of massive sunrise graph~\cite{Remiddi:2016gno,Adams:2016xah}, see eq.~\eqref{eqn:roots}.
In order to express the kite in the region $0<z<1$, we choose the following ordering of the four roots
\begin{equation}
\vec{a} = \left\{ \frac{1}{2} \left( 1 - \sqrt{1+\rho}\right),  \frac{1}{2} \left( 1 + \sqrt{1+\rho}\right),
\frac{1}{2} \left( 1 - \sqrt{1+\bar{\rho}}\right),  \frac{1}{2} \left( 1 + \sqrt{1+\bar{\rho}}\right)\right\}\,.
\end{equation}
This ensures that by giving a positive imaginary part to $z$ we have $0 \leq \lambda \leq 1$, with $\lambda$ defined in eq.~\eqref{eq:lambda4}.
With this choice the first period $\omega_1$ of the elliptic curve is purely real while the second period $\omega_2$ is purely imaginary.
 
We can now easily compute the kite with equal masses in terms of pure eMPLs.
The integral is finite in $D=4$ and by defining
\beq
K(p^2,m^2) = \frac{1}{m^4} 
\frac{1}{z}\left[ K_0(z) + \mathcal{O}(\epsilon) \right]
\eeq
we find for the first order in the $\epsilon$-expansion
\begin{align}
K_0(z) &= 
\frac{1}{6} \Big[-9 \cEf{-1 & -1 & 1}{0 & 0 & 1 }{1}{\vec{a}}
-9 \cEf{-1 & -1 & 1 }{0 & 1 & 1 }{1}{\vec{a}}
-18 \cEf{-1 & -1 & 1 }{0 & \infty  & 1 }{1}{\vec{a}}
-9 \cEf{-1 & 1 & -1 }{0 & 1 & 0 }{1}{\vec{a}} \nonumber \\&
-9 \cEf{-1 & 1 & -1 }{0 & 1 & 1 }{1}{\vec{a}}
-18 \cEf{-1 & 1 & -1 }{0 & 1 & \infty  }{1}{\vec{a}}
+3 \cEf{-1 & -1 & 1 }{\infty  & 0 & 1 }{1}{\vec{a}}
+3 \cEf{ -1 & -1 & 1 }{\infty  & 1 & 1 }{1}{\vec{a}} \nonumber \\&
+6 \cEf{-1 & -1 & 1 }{\infty  & \infty  & 1 }{1}{\vec{a}}
+3 \cEf{-1 & 1 & -1 }{\infty  & 1 & 0 }{1}{\vec{a}}
+3 \cEf{-1 & 1 & -1 }{\infty  & 1 & 1 }{1}{\vec{a}}
+6 \cEf{-1 & 1 & -1 }{\infty  & 1 & \infty  }{1}{\vec{a}} \nonumber \\&
+3 \cEf{1 & -1 & -1 }{0 & 0 & 0 }{1}{\vec{a}}
+3 \cEf{1 & -1 & -1 }{0 & 0 & 1 }{1}{\vec{a}}
+6 \cEf{1 & -1 & -1 }{0 & 0 & \infty  }{1}{\vec{a}}
-6 \cEf{1 & -1 & -1 }{0 & 1 & 0 }{1}{\vec{a}} \nonumber \\&
-6 \cEf{1 & -1 & -1 }{0 & 1 & 1 }{1}{\vec{a}}
-12 \cEf{1 & -1 & -1 }{0 & 1 & \infty  }{1}{\vec{a}}
-3 \cEf{1 & -1 & -1 }{0 & \infty  & 0 }{1}{\vec{a}}
-3 \cEf{1 & -1 & -1 }{0 & \infty  & 1 }{1}{\vec{a}} \nonumber \\&
-6 \cEf{1 & -1 & -1 }{0 & \infty  & \infty  }{1}{\vec{a}}
-6 \cEf{1 & -1 & -1 }{\xi & 0 & 0 }{1}{\vec{a}}
-6 \cEf{1 & -1 & -1 }{ \xi & 0 & 1 }{1}{\vec{a}}
-12 \cEf{1 & -1 & -1 }{ \xi & 0 & \infty  }{1}{\vec{a}} \nonumber \\&
+3 \cEf{ 1 & -1 & -1 }{ \xi & 1 & 0 }{1}{\vec{a}} 
+3 \cEf{1 & -1 & -1 }{\xi & 1 & 1 }{1}{\vec{a}}
+6 \cEf{1 & -1 & -1 }{\xi & 1 & \infty  }{1}{\vec{a}} \nonumber \\&
+9 \cEf{1 & -1 & -1 }{\xi & \infty  & 0 }{1}{\vec{a}}
+9 \cEf{1 & -1 & -1 }{ \xi & \infty  & 1 }{1}{\vec{a}} 
+18 \cEf{1 & -1 & -1 }{ \xi & \infty  & \infty  }{1}{\vec{a}}\nonumber \\& 
+6 \cEf{1 & 1 & 1 }{\xi & 0 & 0 }{1}{\vec{a}} 
-2 \pi ^2 G(0;z)+ \pi ^2 G(1;z) 
-3 G(0,0,0;z)+6 G(0,1,0;z) \nonumber \\& -12 G(0,1,1;z)+3 G(1,0,0;z) +6 G(1,0,1;z)+27 \zeta_3\Big]\nonumber \\&
+ 2\pi i \Big[2 
\cEf{1 & 0 & -1}{ \xi & 0 & \infty  }{1}{\vec{a}}
 +\cEf{1 & 0 & -1}{\xi & 0 & 0 }{1}{\vec{a}}
 +\cEf{1 & 0 & -1}{ \xi & 0 & 1 }{1}{\vec{a}}
 +2 \cEf{0 & -1 & 1}{ 0 & \infty  & 1 }{1}{\vec{a}} \nonumber \\& \qquad  \;\;\,
 +2 \cEf{0 & 1 & -1}{ 0 & 1 & \infty  }{1}{\vec{a}}
 +\cEf{0 & -1 & 1}{ 0 & 0 & 1 }{1}{\vec{a}}
 +\cEf{0 & -1 & 1}{ 0 & 1 & 1 }{1}{\vec{a}}
 +\cEf{0 & 1 & -1}{ 0 & 1 & 0 }{1}{\vec{a}}\nonumber \\&  \qquad  \;\;\,
 +\cEf{0 & 1 & -1}{ 0 & 1 & 1 }{1}{\vec{a}} \Big]\,, \label{eq:kitecE4}
\end{align}
where we defined $\xi = 1/(1-z)$.
As it is easy to see, the result is expressed in terms of MPLs and eMPLs of uniform weight three.
We stress that, in spite of the explicit imaginary parts in eq.~\eqref{eq:kitecE4}, the 
result is real for $0<z<1$. We have checked eq.~\eqref{eq:kitecE4} agrees numerically with the Feynman parameter representation for the kite integral.
We observe that $z_{\xi}$ is not a rational point, so individual eMPLs in eq.~\eqref{eq:kitecE4} cannot be expressed in terms of iterated integrals of modular forms, even though 
it is known that this is the case for the kite integral~\cite{Adams:2017ejb}. We stress that this is not a contradiction, and the $\xi$-dependence can cancel in the combination in eq.~\eqref{eq:kitecE4}.
Since the purpose of this paper is only to show that the kite integral evaluates to a pure function of weight three, we do not investigate this further.


\subsection{Elliptic two-loop three-point functions}
\label{sec:pure_triangle}

In this section we consider a three-point function with a massive closed loop in $D=4-2\eps$ dimensions (see fig.~\ref{fig:examples}d),
\begin{align}
T&(q^2,m^2) =  \\
\nonumber&\!\! -\frac{e^{2\gamma_E\eps}}{\pi^D}\!\!\int\!\!\frac{d^Dk\,d^Dl}{(k-p_1)^2((l-p_1)^2-m^2)(k+p_2)^2((k-l+p_2)^2-m^2)((k-l)^2-m^2)(l^2-m^2)}\,,
\end{align}
with $$p_1^2 = p_2^2 = 0\,, \quad (p_1+p_2)^2 = q^2\,.$$
 This integral contributes to $t\bar{t}$ production at two loops, as well as to two-loop processes like the production of a pair of photons or jets, or a massive weak or Higgs boson in association with a jet. It was computed for the first time using the differential equations technique in ref.~\cite{vonManteuffel:2017hms}, where it was expressed in terms of iterated integrals over complete elliptic integrals. We now show that this integral can be expressed in terms of eMPLs in a natural way. The details of the computation will be presented elsewhere~\cite{trianglepaper}, while here we only present the final result. 
 Seen as a function of $q^2$, $T(q^2,m^2)$ develops a discontinuity as $q^2 \geq 0$ and as $q^2 \geq 4m^2$.
For the scope of this paper, we limit ourselves to consider the Euclidean region where $q^2<0$, and we  find
\beq
T(q^2,m^2) = \frac{32\,\omega_1}{q^4(1+\sqrt{1-16a})}\,\left[T_0(a) + 3\,T_-(a) + 5\,T_+(a)+\ord(\eps)\right]\,,
\eeq
with $a=m^2/(-q^2)$ and 
\begin{align}
T_0(a) &\,= \frac{1}{2} \left(\zeta _2-\log ^2a\right) \cEf{0&-1}{0&\infty }{1}{\vec b}+\log a \,\left[\cEf{0&-1&1}{0&\infty &0}{1}{\vec b}+\cEf{0&-1&1}{0&\infty &1}{1}{\vec b}\right]\\
\nonumber&\,-\cEf{0&-1&1&1}{0&\infty &0&0}{1}{\vec b}-\cEf{0&-1&1&1}{0&\infty &0&1}{1}{\vec b}-\cEf{0&-1&1&1}{0&\infty &1&0}{1}{\vec b}-\cEf{0&-1&1&1}{0&\infty &1&1}{1}{\vec b}\,,\\
\nonumber T_-(a) &\,= \zeta _2\, \cEf{-1&0}{\infty &0}{r_-}{\vec b}+\cEf{-1&0&1&1}{\infty &0&0&0}{r_-}{\vec b}+\cEf{-1&0&1&1}{\infty &0&0&1}{r_-}{\vec b}-\cEf{-1&0&1&1}{\infty &0&1&0}{r_-}{\vec b}\\
\nonumber&\,-\cEf{-1&0&1&1}{\infty &0&1&1}{r_-}{\vec b}+\cEf{-1&1&0&1}{\infty &0&0&1}{r_-}{\vec b}-\cEf{-1&1&0&1}{\infty &1&0&0}{r_-}{\vec b}+\cEf{1&-1&0&1}{0&\infty &0&1}{r_-}{\vec b}\\
\nonumber&\,-\cEf{1&-1&0&1}{1&\infty &0&0}{r_-}{\vec b}+\log (1-r_-) \,\cEf{-1&0&1}{\infty &0&0}{r_-}{\vec b}-\log r_-\, \cEf{-1&0&1}{\infty &0&1}{r_-}{\vec b}\\
\nonumber&\,-\frac{3}{2} \zeta _2\, \cEf{-1}{\infty }{r_-}{\vec b}\,,\\
\nonumber T_+(a) &\,= \cEf{1&-1&0&1}{0&\infty &0&1}{r_+}{\vec b}-\cEf{1&-1&0&1}{0&\infty &0&0}{r_+}{\vec b}-\cEf{1&-1&0&1}{1&\infty &0&0}{r_+}{\vec b}+\cEf{1&-1&0&1}{1&\infty &0&1}{r_+}{\vec b}\\
\nonumber&\,+\frac{i \pi }{4} \left[\cEf{1&-1}{0&\infty }{r_+}{\vec b}+\cEf{1&-1}{1&\infty }{r_+}{\vec b}-4 \cEf{1&-1&0}{0&\infty &0}{r_+}{\vec b}-4 \cEf{1&-1&0}{1&\infty &0}{r_+}{\vec b}\right]\,.
\end{align}
The vector of branch points is
\beq
\vec b = \left(0, \frac{1}{2}(1-\sqrt{1-16a}), \frac{1}{2}(1+\sqrt{1-16a}),1\right)\,,
\eeq
and we have introduced the shorthands
\beq
r_{\pm} = \frac{1}{2}(1-\sqrt{1\pm 4a})\,.
\eeq
We see that the functions $T_0$ and $T_{\pm}$ have uniform weight three. Recalling that we had assigned weight one to the period $\omega_1$, we see that $T$ has uniform weight four, at least for the leading term in the $\eps$ expansion, though we believe that this holds in general. This agrees with the case of massless propagators, which is known to give rise to a function of uniform weight four~\cite{Gonsalves:1983nq,vanNeerven:1985xr,Kramer:1986sr,Gehrmann:2005pd}.



\section{Pure building blocks}
\label{sec:pure_building_blocks}

The aim of this section is to provide a concise summary of the length and weight of the different building blocks of uniform weight that we have encountered in our work, together with the motivation why this weight or length is assigned to a given object. 

\subsection{The length of a period}
We start by analysing the length of a period. In the case of MPLs, both ordinary and elliptic, we have defined the length as the number of integrations. Here we present an alternative definition which seems more widely applicable and reduces to the naive definition as the number of iterated integrations in the case of MPLs and eMPLs. Loosely speaking, the length is defined as the minimal number of iterated differentiations needed to annihilate a given unipotent period. The discussion is inspired by, and follows closely, the construction of the coradical filtration in Section 2.5 of ref.~\cite{Brown:coaction}. 

We start by recalling that we can define a coaction on unipotent periods. We only present the main points, and we refer to ref.~\cite{Brown:coaction,Broedel:2018iwv} for a more detailed exposition. Let us consider a vector $(U_1,\ldots,U_p)^T$ of unipotent periods. By definition, it satisfies a unipotent differential equation, i.e., a differential equation without homogeneous term. In other words, we can write
\beq
dU_{i} = \sum_{j=1}^pA_{ij} U_j\,,
\eeq
where $(A_{ij})$ is a nilpotent matrix of one-forms (at this point we do not restrict ourselves to one-forms with logarithmic singularities). Without loss of generality we can assume the matrix strictly upper triangular. Then the coaction can be defined recursively by~\cite{Brown:coaction}
\beq
\Delta(U_i) = U_i\otimes 1 + \sum_{j=1}^p \left[\Delta(U_j)\big| A_{ij}\right]\,.
\eeq
Loosely speaking, the coaction encodes all iterated differentials of $(U_1,\ldots,U_p)^T$. We refer to ref.~\cite{Broedel:2018iwv} for a more detailed introduction to this topic. Note that since the differential of a constant is zero, the coaction acts trivially on constants, 
\beq
\Delta(c) = c\otimes 1\,,\quad \textrm{ $c$ constant}\,.
\eeq
We extend the coaction so that it also acts trivially on semi-simple periods.

Next, let us define $\Delta'\equiv\Delta-\textrm{id}\otimes 1$. It is easy to see that $\Delta'$ annihilates all objects on which $\Delta$ acts trivially, and more generally we have
\beq
\Delta'(U_i) = \sum_{j=1}^p \left[\Delta(U_j)\big| A_{ij}\right]\,.
\eeq
Let us define $\Delta'_{r+1} \equiv (\Delta'\otimes 1)(\Delta'_r\otimes \textrm{id})$, and the recursion starts with $\Delta'_1=\Delta'$. It is easy to see that for every period $x$ there is a smallest integer $k$ such that $\Delta'_{k+1}(x)=0$. We call this smallest integer the length $k$ of $x$. Since the coaction was defined by means of the total differential, we can easily see that the length $k$ of a unipotent period is the smallest integer such that the unipotent period is annihilated by a $(k+1)$-fold differential. 

This definition of length agrees with the naive notion of length as the number of integrations in the case of iterated integrals. Indeed, every time we act with a differential, we remove one integration, and so if we act with more differentials than we have integrations, we obtain zero. The advantage of our more involved definition of length is that it applies more generally to all periods, not just those defined as iterated integrals. In particular, if $x$ is either semi-simple or constant, then $\Delta'(x)=0$, and so every semi-simple or constant period has length zero. Moreover, we can easily see that the length of a product of two periods of length $k_1$ and $k_2$ is the sum of their lengths, $k_1+k_2$. Finally, let us consider the modular parameter $\tau$. We have seen in Section~\ref{subsec:empls} that $\tau$ is unipotent, and we have~\cite{Broedel:2018iwv}
\beq
\Delta(\tau) = \tau \otimes 1+1\otimes [d\tau]\,.
\eeq
Hence we have $\Delta'(\tau) = 1\otimes [d\tau]$ and $\Delta'_2(\tau) = \Delta'(1)\otimes[\tau] = 0$, and so $\tau$ is a unipotent period of length one. Note that this agrees with alternative representations of $\tau$, e.g.,
\beq\label{eq:tau_length}
\tau = \frac{\log q}{2\pi i} = \gamtt{0}{0}{\tau}{\tau} = I\left(\begin{smallmatrix} 0& 0\\ 0& 0\end{smallmatrix};\tau\right)\,,\qquad q=e^{2\pi i\tau}\,.
\eeq
All the quantities appearing in these identities have length one.

Note that even though the length of an $k$-fold iterated integral is generically $k$, there can be special instances where we have a `length drop', e.g., when the iterated integral is evaluated at some special point where it evaluates to a constant (which has length zero). For example, the length of $\gamtt{0}{0}{z}{\tau} = z$ is one, because $\Delta'(\gamtt{0}{0}{z}{\tau}) = 1\otimes [dz]$ and $\Delta'_2(\gamtt{0}{0}{z}{\tau}) = 0$. However, if we evaluate $\gamtt{0}{0}{z}{\tau}$ at some constant value for $z$, say $z=1$, then the length of $\gamtt{0}{0}{z}{\tau} = \gamtt{0}{0}{1}{\tau} = 1$ drops to zero.

\subsection{Building blocks of uniform weight}

After our discussion of the length of a period, let us turn to the weight. While in the case of the length we could give a fully general definition valid for arbitrary periods, our discussion of the weight will be based mostly on empirical observations. Beyond ordinary MPLs we \emph{cannot} follow the folklore in the physics literature and define the weight through the action of the differential. Indeed, extending this folklore definition from ordinary MPLs to eMPLs would not lead to the weight, but to the length discussed in the previous section. In the case of ordinary MPLs, the weight is always equal to the length (except in the case of $i\pi$, which has weight one and length zero), and so the notions of length and weight are indistinguishable. Beyond ordinary MPLs, however, the two concepts are no longer identical. In the following we present a definition of weight that extends the `transcendental weight' in the physics literature to the elliptic case and that seems to be consistent with the idea that certain Feynman integrals should evaluate to functions of `uniform transcendental weight'. We emphasise that this definition is purely based on empirical observations. In particular, we do not know if or how this notion of (transcendental) weight is connected to the weight filtration defined on motivic periods in the mathematics literature (cf., e.g., ref.~\cite{Brown:coaction}). It would be interesting to clarify this point in the future.

Our definition of weight is summarised in Table~\ref{tab:summary}. We postulate that the weight is additive, i.e., the weight of a product of two quantities of weight $n_1$ and $n_2$ is $n_1+n_2$. Most of the entries in this table have already been discussed in previous sections. In the remainder of this section we only focus on those entries which have not appeared in this paper so far.

First, we have already seen in Section~\ref{sec:pure_sunrise} that it is natural to assign weight one to $\omega_1$. From eq.~\eqref{eq:tau_length} we see that the only consistent weight we can assign to $\tau$ is zero. As a consequence, $\omega_2 = \omega_1\tau$ has weight one.

Next, let us discuss the weight of modular forms. We assign weight zero to a modular form whose Fourier coefficients are all algebraic. The same definition extends to the case of meromorphic and quasi-modular forms. Note that the case of algebraic Fourier coefficients is not special but rather generic, and one can often show that there is a basis with this property for a given vector space of modular forms. From eq.~\eqref{eq:g_q_exp} we see that the $q$-expansions of $g^{(n)}(z,\tau)$ and $h^{(n)}_{N,r,s}(\tau)$ involve additional powers of $(2\pi i)^n$ at each order, and we therefore assign weight $n$ to $g^{(n)}(z,\tau)$ and $h^{(n)}_{N,r,s}(\tau)$. As a consequence, we see from eqs.~\eqref{eq:Z4_to_g1} and~\eqref{eq:G_infty_def} that $Z_4(x,\vec a)$ and $G_{\ast}(\vec a)$ have weight zero. 

Finally, let us comment on the weight of the quasi-periods. We know from eq.~\eqref{eq:quasi-periods4} that $\eta_i$ only depends on the cross ratio $\lambda$. One can show that if 
\beq
\tau = \frac{\omega_2}{\omega_1} = i\frac{\textrm{K}(1-\ell)}{\textrm{K}(\ell)}\,,
\eeq
then the inverse is $\ell = \lambda(\tau)$, where $\lambda$ is the modular $\lambda$ function. We can thus see $\eta_i$ equally well as a function of only $\tau$. The same holds of course for $\omega_i$. We have~\cite{Zemel}
\beq
\eta_1(\lambda(\tau)) = -\frac{h^{(2)}_{1,0,0}(\tau)}{2\omega_1(\lambda(\tau))}\,.
\eeq
We need to assign weight one to $\eta_1$. $\eta_2$ instead does not have uniform weight, as can be seen from the Legendre relation in eq.~\eqref{eq:Legendre} and the fact that $\omega_1$, $\omega_2$ and $\eta_1$ have weight one. We note that this assignment for the weight is tightly connected to our choice of the period matrix of the elliptic curve, and its factorisation into a semi-simple and a unipotent matrix, cf.~eqs.~\eqref{eq:per_mat} and~\eqref{eq:US_matrix}.

Let us conclude this section with an empirical observation. In the course of our computation of the integrals shown in fig.~\ref{fig:examples}~\cite{trianglepaper,Bhabhapaper}, we have obtained functional relations among eMPLs and ordinary MPLs. In all cases we find that the weight is preserved by relations relating eMPLs with different arguments. For example, consider the following function
\beq\label{eq:kite_log}
f(x) = \log\frac{x[\mu_1+(1-\mu_3)(1-x)]+\mu_2(1-x) -(1+\mu_3)y}{x[\mu_1+(1-\mu_3)(1-x)]+\mu_2(1-x) +(1+\mu_3)y}\,.
\eeq
The variable $y$ is not an independent variable, but it is constrained by the quartic polynomial equation
\beq\label{eq:kite_curve}
y^2 = x^4 + c_3\,x^3+c_2\,x^2+c_1\,x+c_0 = (x-a_1)(x-a_2)(x-a_3)(x-a_4)\,.
\eeq
The coefficients appearing inside the quartic polynomial are
\beq\bsp
c_3&\,=\frac{2 \left(-\mu _3^2+\mu _1 \mu _3-\mu _2 \mu _3+2 \mu _3+\mu _1-\mu _2-1\right)}{\left(\mu _3-1\right)^2}\,\\
c_2 &\, =\frac{\mu _1^2-2 \mu _2 \mu _1-2 \mu _3 \mu _1-2 \mu _1+\mu _2^2+\mu _3^2+4 \mu _2+4 \mu _2 \mu _3-2 \mu _3+1}{\left(\mu _3-1\right)^2}\,,\\
c_1&\,=-\frac{2 \mu _2 \left(-\mu _1+\mu _2+\mu _3+1\right)}{\left(\mu _3-1\right)^2}\,,\\
c_0 &\,= \frac{\mu _2^2}{\left(\mu _3-1\right)^2}\,.
\esp\eeq
The roots $a_i$ of the quartic polynomial are complicated algebraic functions whose explicit form is irrelevant for the following. The logarithm in eq.~\eqref{eq:kite_log} appears in the computation of the two-loop kite integral with three different masses~\cite{trianglepaper}, with $\mu_i=m_i^2/p^2>1$ and $x$ is a Feynman parameter that still needs to be integrated over the range $[0,1]$. In order to perform the integral over $x$, it is useful to cast this logarithm in the form of an integral where $x$ only appears as the upper integration limit. Since eq.~\eqref{eq:kite_curve} defines an elliptic curve, it is easy to see that such an integral representation will involve eMPLs. Indeed, we find,
\beq\bsp\label{eq:kite_example}
f(x) &\,= \log\mu_2 + \cEf{-1}{\infty}{x}{\vec{a}} - \cEf{-1}{\frac{\mu_2}{\mu_2-\mu_1}}{x}{\vec{a}} -\cEf{-1}{0}{x}{\vec{a}}\\
&\,-\cEf{-1}{1}{x}{\vec{a}}-4\pi i\, \cEf{0}{0}{x}{\vec{a}}\,.
\esp\eeq
We see that every term in the right-hand side has weight one, just like the logarithm in eq.~\eqref{eq:kite_log}. We stress that it is crucial that the weight is not identified with the number of integrations, because otherwise the last term in eq.~\eqref{eq:kite_example} would not have weight one. We have derived a large variety of identities of this type up to weight three, and in all cases we observe that the weight is conserved. We therefore conjecture that this observation holds in general, and extends the corresponding property for ordinary MPLs.

\begin{table}[!th]
\begin{center}
\begin{tabular}{c|c|c|c}
\hline\hline
Name & Unipotent & Length & Weight \\
\hline \hline
Rational Functions & No & 0 & 0\\
Algebraic Functions & No & 0 & 0\\
\hline
$i \pi$ & No & 0 & 1\\
$\zeta_{2n}$ & No & 0 & $2n$\\
$\zeta_{2n+1}$ & Yes & $0$ & $2n+1$\\
$\log x$ & Yes & 1 & 1\\
$\textrm{Li}_n(x)$ & Yes & $n$ & $n$\\
$G(c_1,\ldots,c_k;x)$ & Yes & $k$ & $k$\\
\hline
$\omega_1$ & No & $0$ &$1$\\
$\eta_1$ & No & $0$ &$1$\\
$\tau$ & Yes & $1$ &$0$\\
$g^{(n)}(z,\tau)$ & No & $0$ &$n$\\
$h^{(n)}_{N,r,s}(\tau)$ & No & $0$ &$n$\\
$Z_4(c,\vec a)$ & No & 0 & 0\\
$G_{\ast}(\vec a)$ & No & 0 & 0\\
$\cEf{n_1&\ldots&n_k}{c_1&\ldots&c_k}{x}{\vec a}$ & Yes & $ k$ & $\sum_i|n_i|$\\
$\gamtt{n_1&\ldots&n_k}{z_1&\ldots&z_k}{z}{\tau}$ & Yes & $ k$ & $\sum_in_i$\\
$I\left(\begin{smallmatrix} n_1& N_1\\ r_1& s_1\end{smallmatrix}\big|\ldots\big|\begin{smallmatrix} n_k& N_k\\ r_k& s_k\end{smallmatrix};\tau\right)$ & Yes & $k$ &  $\sum_in_i$\\
\hline\hline
\end{tabular}
\caption{\label{tab:summary}Weight and length of the different building blocks encountered when working with elliptic Feynman integrals. Note that these assignments are connected to our choice of the factorisation of the period matrix of the elliptic curve into a semi-simple and a unipotent matrix, cf.~eqs.~\eqref{eq:per_mat} and~\eqref{eq:US_matrix}.}
\end{center}
\end{table}


\section{Conclusion}
\label{sec:conclusions}

In this paper we have introduced a generalisation of the notion of pure functions that goes beyond the case of ordinary MPLs studied in the literature~\cite{ArkaniHamed:2010gh,Henn:2013pwa}. This definition of purity applies at the same time to ordinary MPLs and to the eMPLs introduced in the mathematics literature~\cite{BrownLevin}. 

In Section~\ref{sec:applications} we have illustrated the use of these pure eMPLs in the context of elliptic Feynman integrals. We have studied analytic results for elliptic Feynman integrals with up to four external legs. If one (or more) of the scales vanish, the integrals can be expressed in terms of ordinary MPLs of uniform weight. In all cases we observe that this weight agrees with the weight of the eMPLs in the elliptic case. This is the first time that a notion of uniform weight is observed in the context of Feynman integrals that evaluate to eMPLs. Given the important role played by pure functions of uniform weight for non-elliptic Feynman integrals, we believe that our findings will have an impact on future studies of elliptic Feynman integrals, both for practical computations and for our understanding of the mathematics of multi-loop integrals and perturbative scattering amplitudes in general.

Let us conclude this paper by commenting on possible implications of our work for scattering amplitudes in the $\cN=4$ Super Yang-Mills (SYM) theory. It is known that there is a specific component of the two-loop $10$-point N$^3$MHV super-amplitude which is equal to a double-box integral which cannot be expressed in terms of ordinary polylogarithms~\cite{CaronHuot:2012ab,Nandan:2013ip,Bourjaily:2017bsb}. In ref.~\cite{Bourjaily:2017bsb} this double-box integral was written as a one-fold integral, which can schematically be represented as (cf. eq.~(2) of ref.~\cite{Bourjaily:2017bsb}),
\beq
I_{\textrm{db}}^{\textrm{ell}} \sim \int \frac{d\alpha}{\sqrt{Q(\alpha)}}\,\cG_3(\alpha)\,,
\eeq
where $Q(\alpha)$ denotes a quartic polynomial in $\alpha$ whose coefficients depend on the dual conformally invariant cross ratios, and $\cG_3(\alpha)$ denotes a pure combination of MPLs of weight three. So far it is not known if this integral can be evaluated in terms of eMPLs, because the arguments of the MPLs in $\cG_3(\alpha)$ are algebraic functions of $\alpha$ that involve not only the square root $\sqrt{Q(\alpha)}$, but also additional square roots with a quadratic dependence on $\alpha$. We can, however, use results from this paper to analyse the weight of $I_{\textrm{db}}^{\textrm{ell}}$. We can write
\beq\label{eq:edb}
I_{\textrm{db}}^{\textrm{ell}} \sim \frac{\omega_1}{c_4}\,T_{\textrm{db}}^{\textrm{ell}} \quad\textrm{with}\quad T_{\textrm{db}}^{\textrm{ell}} = \int d\alpha\, \Psi_0(0,\alpha)\,\cG_3(\alpha)\,,
\eeq
where $\omega_1$ denotes one of the periods of the elliptic curve defined by the polynomial equation $\beta^2 = Q(\alpha)$ and $\Psi_0(0,\alpha)$ is defined in eq.~\eqref{eq:pure_psi0}. We see that $T_{\textrm{db}}^{\textrm{ell}}$ defines a pure function of length four and weight three. Since $\omega_1$ has weight one, we conclude that $I_{\textrm{db}}^{\textrm{ell}}$ has uniform weight four. This is in agreement with known results for two-loop amplitudes in $\cN=4$ SYM that evaluate to ordinary MPLs, and hints towards the fascinating possibility of an extension of the principle of uniform transcendentality beyond the the case of ordinary MPLs.

\section*{Acknowledgments}
CD and LT are grateful to the Mainz Institute for Theoretical Physics (MITP) for its hospitality and its partial support during the completion of this work.
This research was supported by the the ERC grant 637019 ``MathAm'', and the U.S.
Department of Energy (DOE) under contract DE-AC02-76SF00515.

\appendix 

\section{Expressing $z_{\ast}$ in terms of elliptic integrals}
\label{app:zstar}

In this appendix we give some details on how to derive eq.~\eqref{eq:zstar}. We start by discussing the case where the branch points $a_i$ are real and ordered according to $a_1<a_2<a_3<a_4$, which is the case considered in the main text. We comment on other cases towards the end of this section, and we show that in those other cases one needs to be careful about signs.

We start from the integral definition for $z_{\ast}$ in eq.~\eqref{eq:zstar_int}. We assume that the branch points are real and ordered, and we follow eq.~\eqref{eq:rsigns} for the choice of the branches of the square root. Then, in the whole integration region we have
\beq\label{eq:rsigns_app_1}
y = \sqrt{P_4(x)} = -\sqrt{|P_4(x)|}\,,\qquad P_4(x) = (x-a_1)\ldots(x-a_4)\,, \quad x<a_1\,,
\eeq
and eq.~\eqref{eq:zstar_int} reduces to
\beq
z_{\ast} = -\frac{\sqrt{a_{31}a_{42}}}{4\textrm{K}(\lambda)}\int_{a_1}^{-\infty}\frac{dx}{\sqrt{|P_4(x)|}}\,.
\eeq
We can perform the change of variables
\begin{align}
t^2 = \frac{\left(a_3-a_1\right) \left(x-a_4\right)}{\left(a_1-a_4\right) \left(a_3-x\right)}\,, \label{eq:chvar}
\end{align}
and we find
\beq\label{eq:zstar_app}
z_{\ast} = -\frac{1}{2\textrm{K}(\lambda)}\int^{\sqrt{\alpha}}_1\frac{dt}{\sqrt{(1-t^2)(1-\lambda t^2)}} = \frac{1}{2}-\frac{\textrm{F}(\sqrt{\alpha}|\lambda)}{2\textrm{K}(\lambda)} = \cZ_{\ast}(\alpha,\lambda)\,,
\eeq
where in the last step we used $\textrm{F}(1|\lambda) = \textrm{K}(\lambda)$. In this way we recover eq.~\eqref{eq:zstar}.

Let us conclude by commenting on the overall sign of eq.~\eqref{eq:zstar_app}. It is easy to see that the overall sign is dictated by our choice for the branch of the square root in the range $x<a_1$, cf. eq.~\eqref{eq:rsigns_app_1}. Indeed, if we had chosen the convention that $\sqrt{P_4(x)} = +\sqrt{|P_4(x)|}$ is real and positive for $x<a_1$, the overall sign of $z_{\ast}$ would be reversed. We thus conclude that the overall sign of $z_{\ast}$ is tightly linked to the (conventional) choice of the branches of the square root. The formula in eq.~\eqref{eq:zstar} is valid for the choice in eq.~\eqref{eq:rsigns} in the case where the branch points are real and ordered in the natural way, but care is needed when applying it to different cases. 

For example, let us discuss what happens in the case where the branch points $a_i$ are pairwise complex conjugate (as is often the case in physics applications). For concreteness we assume that $a_1=a_2^\ast$, $a_3=a_4^\ast$, $\textrm{Re}(a_1)<\textrm{Re}(a_3)$ and $\textrm{Im}(a_2),\textrm{Im}(a_3)>0$ and $\textrm{Im}(a_1),\textrm{Im}(a_4)<0$. We emphasise that some formulas may change if we choose a different configuration. For example, if we had considered the situation where $\textrm{Im}(a_1),\textrm{Im}(a_4)>0$ and $\textrm{Im}(a_2),\textrm{Im}(a_3)<0$ (which is related to the previous one via complex conjugation), the signs of all imaginary parts would change. Next, let us discuss the choice of the branches of the square root. We cannot use eq.~\eqref{eq:rsigns}, because that convention only makes sense when the branch points are real. Instead, since $P_4(x)$ is positive definite for pairwise complex conjugate branch points, it is natural to choose $\sqrt{P_4(x)} = \sqrt{|P_4(x)|} > 0$ everywhere on the real axis. While this convention is very natural for applications, it is easy to see that there is a tension between this choice and the choice in eq.~\eqref{eq:rsigns_app_1} for $x<\textrm{Re}(a_1)$, and so in this case the formula for $z_{\ast}$ has a different overall sign,
\beq\label{eq:app_2}
z_{\ast} = -\cZ_{\ast}(\alpha,\lambda)\,.
\eeq
This explains the sign difference in eq.~\eqref{eq:z*_sunrise} compared to eq.~\eqref{eq:zstar}.

Finally, let us mention that the same reasoning shows that also the sign of eq.~\eqref{eq:G_infty_KE} and eq.~\eqref{eq:G_start_algebraic} depend on the choice of the branches for the square root and the ordering of the branch points. As a general rule, we find that the following equations are independent of the chosen convention for the branches of square root,
\beq\bsp\label{eq:app_1}
G_{\ast}(\vec a) &\,= \frac{1}{\omega_1}\,g^{(1)}(z_{\ast},\tau)\,,\\
\cZ_{\ast}(\alpha,\lambda)&\, = \frac{1}{2}-\frac{\textrm{F}(\sqrt{\alpha}|\lambda)}{2\textrm{K}(\lambda)}\,,\\
\frac{1}{\omega_1}\,g^{(1)}\left(\cZ_{\ast}(\alpha,\lambda),\tau\right)
&\, = \left(\frac{2  \eta_1}{ \omega_1}-\frac{\lambda }{3}+\frac{2}{3}\right) \textrm{F}\!\left(\sqrt{\alpha}|\lambda \right)-\textrm{E}\!\left(\sqrt{\alpha }|\lambda \right)+\sqrt{\frac{\alpha  (\alpha  \lambda -1)}{\alpha -1}}\\
&\,=\frac{(1-\lambda)\left[\lambda\,  \alpha '(\lambda ) + \alpha\right]}{\sqrt{ \alpha (1-\alpha )  (1-\alpha  \lambda )}}
-b\,  \frac{ 2\pi i }{ \omega_1}\,,
\esp\eeq
where the last equality is subject to the conditions discussed at the end of Section~\ref{sec:pure_eMPLs}. Once the overall sign of $z_{\ast}$ has been determined using the reasoning at the beginning of the section, the overall sign of $G_{\ast}(\vec a)$ can be determined from eq.~\eqref{eq:app_1}. Let us illustrate this on two examples:
\begin{enumerate}
\item In the case where the branch points are real and ordered in the natural way, and the branches of the square root are chosen according eq.~\eqref{eq:rsigns}, we know that we have (cf. eq.~\eqref{eq:zstar})\
\beq
z_{\ast} = \cZ_{\ast}(\alpha,\lambda) = \frac{1}{2}-\frac{\textrm{F}(\sqrt{\alpha}|\lambda)}{2\textrm{K}(\lambda)}\,.
\eeq
Equation~\eqref{eq:app_1} then implies
\beq\bsp
G_{\ast}(\vec a) &\,= \frac{1}{\omega_1}\,g^{(1)}(z_{\ast},\tau) = \frac{1}{\omega_1}\,g^{(1)}\left(\cZ_{\ast}(\alpha,\lambda),\tau\right)\\
&\, = \left(\frac{2  \eta_1}{ \omega_1}-\frac{\lambda }{3}+\frac{2}{3}\right) \textrm{F}\!\left(\sqrt{\alpha}|\lambda \right)-\textrm{E}\!\left(\sqrt{\alpha }|\lambda \right)+\sqrt{\frac{\alpha  (\alpha  \lambda -1)}{\alpha -1}}\\
&\,=\frac{(1-\lambda)\left[\lambda\,  \alpha '(\lambda ) + \alpha\right]}{\sqrt{ \alpha (1-\alpha )  (1-\alpha  \lambda )}}
-b\,  \frac{ 2\pi i }{ \omega_1}\,,
\esp\eeq
in agreement with eq.~\eqref{eq:G_infty_KE} and eq.~\eqref{eq:G_start_algebraic}.
\item In the case where the branch points are pairwise complex conjugate to each other and $\sqrt{P_4(x)}>0$ everywhere on the real axis, we have seen above that (cf. eq.~\eqref{eq:app_2})
\beq
z_{\ast} = -\cZ_{\ast}(\alpha,\lambda) = \frac{\textrm{F}(\sqrt{\alpha}|\lambda)}{2\textrm{K}(\lambda)}-\frac{1}{2}\,,
\eeq
and eq.~\eqref{eq:app_1} implies
\beq\bsp
G_{\ast}(\vec a) &\,= \frac{1}{\omega_1}\,g^{(1)}(z_{\ast},\tau) = -\frac{1}{\omega_1}\,g^{(1)}\left(\cZ_{\ast}(\alpha,\lambda),\tau\right)\\
&\, = -\left[\left(\frac{2  \eta_1}{ \omega_1}-\frac{\lambda }{3}+\frac{2}{3}\right) \textrm{F}\!\left(\sqrt{\alpha}|\lambda \right)-\textrm{E}\!\left(\sqrt{\alpha }|\lambda \right)+\sqrt{\frac{\alpha  (\alpha  \lambda -1)}{\alpha -1}}\right]\\
&\,=-\left[\frac{(1-\lambda)\left[\lambda\,  \alpha '(\lambda ) + \alpha\right]}{\sqrt{ \alpha (1-\alpha )  (1-\alpha  \lambda )}}
-b\,  \frac{ 2\pi i }{ \omega_1}\right]\,.
\esp\eeq
\end{enumerate}


\section{Analytic expressions for the integration kernels of weight two}
\label{app:kernels_2}

In this appendix we present the analytic forms of the integration kernels $\Psi_{\pm2}(\infty,x,\vec a)$ and $\Psi_{\pm2}(c,x,\vec a)$, with $c\neq \infty$.

\begin{align}
\Psi_{2}(c,x,\vec a) & = -\frac{1}{2}\, \omega _1\, \psi _2(c,x,\vec{a})+\frac{1}{2}\, \omega _1\, \psi _2\left(a_1,x,\vec{a}\right)+\frac{2 c_4 \left(c-a_1\right)}{a_{24} a_{13}}\,\omega_1\,\psi _{-1}(\infty ,x,\vec{a})\\
\nonumber&+ \left[\frac{1}{4}\,Z_4(c,\vec{a})^2+\frac{\left(a_1-c\right) \left(a_2+a_3+a_4-c\right)}{a_{24} a_{13}}\right]\,\omega_1\,\psi _0(0,x,\vec{a})+2\, \omega _1\, \psi _2(\infty ,x,\vec{a})\\
\nonumber&+\frac{1}{2}\, \omega _1\, Z_4(c,\vec{a})\, \psi _{-1}(c,x,\vec{a})\,,\\
\nonumber\\
\Psi_{-2}(c,x,\vec a) & = -\frac{1}{2} \omega _1 \psi _{-2}(c,x,\vec{a})-\frac{1}{2} \omega _1 Z_4(c,\vec{a}) \psi _1(\infty ,x,\vec{a})+\frac{1}{2} \omega _1 Z_4(c,\vec{a}) \psi _1(c,x,\vec{a})\,,\\
\nonumber\\
\nonumber\Psi_{2}(\infty,x,\vec a) & = 2\, \omega _1\, \psi _2(\infty ,x,\vec{a})+\frac{1}{2}\, \omega _1\, \psi _2\left(a_1,x,\vec{a}\right)+ \left[\frac{4 \,a_1\, c_4  }{a_{13} a_{24}}\,G_{\ast}(\vec{a})+ G_{\ast}(\vec{a})^2+2 \frac{\eta _1}{\omega_1}\right.\\
&\left.+\frac{3 a_1^2+\left(a_2+a_3+a_4\right) a_1-2 \left(a_3 a_4+a_2 a_3+a_2a_4\right) }{3 a_{13} a_{24}}\right]\,\omega_1\,\psi _0(0,x,\vec{a})\\
\nonumber&- \left[G_{\ast}(\vec{a})+\frac{\left(3 a_1-a_2-a_3-a_4\right) }{4 c_4}\right]\,\omega_1\,\psi _{-1}(\infty ,x,\vec{a})\,,\\
\nonumber\\
\Psi_{-2}(\infty,x,\vec a) & = -\frac{1}{2}\, \omega _1\, \psi _{-2}(\infty ,x,\vec{a})+ \left[\frac{a_1}{2 c_4}+ G_{\ast}(\vec{a})\right]\,\omega _1\,\psi _1(\infty ,x,\vec{a})\,.
\end{align}

\bibliography{bib}

\end{document}